\documentclass[twocolumn]{aastex6}

\usepackage[english]{babel}

\usepackage[T1]{fontenc}
\usepackage{ae,aecompl}

\usepackage{fancyhdr}
\usepackage{amsfonts}
\usepackage{amsmath}
\usepackage{amssymb}
\usepackage{layout}
\usepackage{graphicx}
\usepackage{multirow}
\usepackage{color}
\usepackage{multirow}

\usepackage{times}
\usepackage{natbib}
\def\be{\begin{equation}}
\def\ee{\end{equation}}

\newif\ifAMStwofonts
\AMStwofontstrue

\graphicspath{{./fig/}}

\shorttitle{PADE parameterisations}
\shortauthors{Rezaei et al.}

\begin{document}

\title{Constraints to dark energy using  PADE parameterisations}
\author{M. Rezaei \altaffilmark{1}}
\author{ M. Malekjani \altaffilmark{1}}
\author{ S. Basilakos  \altaffilmark{2}}
\author{A. Mehrabi \altaffilmark{1}}
\author{D. F. Mota\altaffilmark{3}}

\affil{\altaffilmark{1} Department of Physics, Bu-Ali Sina University, Hamedan 65178, Iran }

\affil{ \altaffilmark{2}Academy of Athens, Research Center for Astronomy and Applied Mathematics, Soranou Efessiou 4, 11-527 Athens, Greece}

\affil{\altaffilmark{3}Institute of Theoretical Astrophysics, University of Oslo, 0315 Oslo, Norway}


\begin{abstract}
We put constraints on dark energy properties using the PADE parameterisation, and compare it to the same constraints using Chevalier-Polarski-Linder (CPL) and $\Lambda$CDM,
at both the background and the perturbation levels. 
The dark energy equation of state parameter of the models 
is derived following the mathematical treatment of PADE expansion.
Unlike CPL parameterisation, the PADE approximation provides
different forms of the equation of state parameter 
which avoid the divergence in the far future.
Initially, we perform a likelihood analysis in order to
put constraints on the model parameters using solely background expansion data
and we find that all parameterisations are consistent with each other.
Then, combining the expansion and the growth rate data 
we test the viability of PADE parameterisations and compare them 
with CPL and $\Lambda$CDM models respectively.
Specifically, we find that the growth rate of 
the current PADE parameterisations is lower than $\Lambda$CDM model at low redshifts, 
while the differences among the models are negligible at high redshifts. 
In this context, we provide for the first time growth index of linear matter
perturbations in PADE cosmologies. 
Considering that dark energy is homogeneous
we recover the well known asymptotic value of the growth index, namely  
$\gamma_{\infty}=\frac{3(w_{\infty}-1)}{6w_{\infty}-5}$, while
in the case of clustered dark energy we obtain 
$\gamma_{\infty}\simeq \frac{3w_{\infty}(3w_{\infty}-5)}{(6w_{\infty}-5)(3w_{\infty}-1)}$.
Finally, we generalize the growth index analysis in the case where $\gamma$
is allowed to vary with redshift and we find that the form of $\gamma(z)$ in PADE parameterisation
extends that of the CPL and $\Lambda$CDM cosmologies respectively.

\end{abstract}


\section{Introduction}
Various independent cosmic observations including those of type Ia
supernova (SnIa) \citep{Riess1998,Perlmutter1999,Kowalski2008},
cosmic microwave background (CMB)
\citep{Komatsu2009,Jarosik:2010iu,Komatsu2011,Ade:2015yua},
large scale structure (LSS), baryonic acoustic oscillation (BAO)
\citep{Tegmark:2003ud,Cole:2005sx,Eisenstein:2005su,Percival2010,Blake:2011rj,Reid:2012sw},
high redshift galaxies \citep{Alcaniz:2003qy}, high redshift galaxy
clusters \citep{Wang1998,Allen:2004cd} and weak gravitational
lensing \citep{Benjamin:2007ys,Amendola:2007rr,Fu:2007qq} reveal that the present universe experiences an accelerated expansion. Within the framework of General Relativity (GR), the physical origin of the cosmic acceleration can be described by invoking the existence of an exotic fluid with sufficiently negative pressure, the so-called Dark Energy (DE). One possibility is that DE consists of the vacuum energy
or cosmological constant $\Lambda$ with constant EoS parameter $w_{\rm \Lambda}=-1$  \citep{Peebles2003}.
Alternatively, the fine-tuning and cosmic coincidence problems \citep{Weinberg1989,Sahni:1999gb,Carroll2001,Padmanabhan2003,Copeland:2006wr} 
led the scientific community to suggest a time evolving energy density with negative pressure. In those models, the EoS parameter is a function
of redshift, $w(z)$ \citep{Caldwell:1997ii,Erickson:2001bq,Armendariz2001,Caldwell2002,Padmanabhan2002,Elizalde:2004mq}. 
A  precise measurement of EoS parameter and its variation with cosmic time 
can provide important clues about the dynamical behavior of
DE and its nature \citep{Copeland:2006wr,Frieman:2008sn,Weinberg:2012es,Amendola:2012ys}.

One possible way to study the  EoS parameter of dynamical DE models is via a parameterisation. 
In literature, one can find many different EoS parameterisations. One of the simplest and earliest parameterisations is the Taylor
expansion of $w_{\rm de}(z)$ with respect to redshift $z$ up to first order as: $w_{\rm de}(z)=w_0+w_{1}z$ \citep{Maor:2000jy,Riess:2004nr}. 
It can also be generalized by considering the second order approximation in Taylor series as: $w_{\rm de}(z)=w_0+w_{1}z+w_{2}z^2$ \citep{Bassett:2007aw}. 
However, these two parameterisations diverge at high redshifts. Hence the well-known Chevallier-Polarski-Linder (CPL) parameterisation, $w_{\rm de}(z)=w_0+w_1(1-a)=w_0+w_1z/(1+z)$, was proposed \citep{Chevallier:2000qy,Linder:2002et}. The CPL parameterisation can be considered as a Taylor
series with respect to $(1-a)$ and was extended to more general case by assuming the second order approximation as:
$w_{\rm de}(a)=w_0+w_1(1-a)+w_2(1-a)^2$ \citep{Seljak:2004xh}. In addition to CPL formula, some purely phenomenological parameterisations have been proposed more recently. For example $w_{\rm de}(z)=w_0+w_1z/(1+z)^{\alpha}$, where $\alpha$ is fixed to 2 \citep{Jassal:2004ej}. In this class, the power law $w_{\rm de}(a)=w_0+w_1(1-a^{\beta})/\beta$ \citep{Barboza:2009ks} and logarithmic $w_{\rm de}(a)=w_0+w_1 \ln{a}$ \citep{Efstathiou:1999tm} parameterisations have been investigated. Another logarithm parameterisation is $w_{\rm de}(z)=w_0/[1+b\ln{(1+z)}]^{\alpha}$, where $\alpha$ is taken to be $1$ or $2$ \citep{Wetterich:2004pv}. Notice that although the CPL is a well-behaved parameterisation at early ($a\ll 1$) and present ($a\sim 1$) epochs, it diverges when the scale factor goes to infinity. 
This is also a common difficulty for the above phenomenological parameterisations. Recently to solve the divergence, several phenomenological 
parameterisations have been introduced \citep[see][ for more details] {Dent:2008ek,Frampton:2011rq,Feng:2012gf}. Notice that most of these EoS 
parameterisations are \textit{ad hoc} and purely written by hand where there is no mathematical principle or fundamental physics behind them. In this work we focus on the PADE parameterisation ( see section \ref{sect:pade}), which from the mathematical point of view seems to be more stable: it does not diverge and can be employed at both small and high redshifts.  
Using the different types of PADE parameterisations to express the EoS parameter of DE $w_{\rm de}$ in terms of redshift $z$, we study the growth of perturbations in the universe.

DE not only accelerate the expansion rate of universe but also change the evolution of 
growth rate of matter perturbations and consequently the formation epochs of large scale structures of 
universe \citep{ArmendarizPicon:1999rj,Garriga:1999vw,ArmendarizPicon:2000dh,Tegmark:2003ud,Pace2010,Akhoury:2011hr}. 
Moreover, the growth of cosmic structures are also affected by perturbations of DE when we deal with dynamical 
DE models with time varying EoS parameter $w_{\rm de}\neq -1$ \citep{Erickson:2001bq,Bean:2003fb,Hu:2004yd,Basilakos:2006us,Ballesteros:2008qk,Basilakos:2009mz,mota1,mota2,mota3,Basilakos:2010fb,Sapone:2012nh,Batista:2013oca,Dossett:2013npa,Basse:2013zua,Pace:2013pea,Batista:2014uoa,Basilakos:2014yda,Pace:2014taa,Nesseris:2014mfa,Mehrabi:2014ema,Mehrabi:2015hva,Malekjani:2015pza,Mehrabi:2015kta,Malekjani:2016edh}.

In addition to the background geometrical data the data coming from the formation of large scale structures provide a valuable
information about the nature of DE. In particular, we can setup a more general formalism in which the background expansion data including 
SnIa, BAO, CMB shift parameter, Hubble expansion data joined with the
growth rate data of large scale structures in order to put constraints on the parameters of cosmology and DE models \citep[see][]{Cooray:2003hd,Corasaniti:2005pq,Basilakos:2010fb,Blake:2011rj,Nesseris:2011pc,Basilakos:2012uu,Yang:2013hra,mota1,mota2,mota3,mota4,mota5,mota6,Contreras:2013bol,Chuang:2013hya,Li:2014mua,Basilakos:2014yda,Mehrabi:2015kta,Mehrabi:2015hva,Basilakos:2016xob,mota7,Malekjani:2016edh,Fay:2016yow,Rivera:2016zzr}.

In this work, following the lines of the above studies and using the latest observational data including the geometrical data set (SnIa, BAO, CMB, big bang nucleosynthesis (BBN), $H(z)$) combined with growth rate data $f(z)\sigma_8$, we perform an overall likelihood statistical analysis to place constraints and find best fit values of the cosmological parameters where the EoS parameter of DE is approximated by PADE parameterisations. Previously, the PADE parameterisations have been studied from different observational tests in Cosmology. For example  in \cite{Gruber:2013wua}, the cosmography analysis has been investigated using the PADE approximation. 
In \cite{Wei:2013jya}, the authors proposed several parameterisations for EoS of DE on the basis of PADE approximation. Confronting these EoS parameterisations with the latest geometrical data, they found that the PADE parameterisations can work well
\citep[for similar studies, see][]{Aviles:2014rma,Zaninetti:2016fju,Zhou:2016nik}. 
Here, for the first time, we study the 
growth of perturbations in PADE cosmologies. After introducing the main ingredients of PADE parameterisations in Sect.\ref{sect:pade}, we study the 
background evolution of the universe 
in Sect.(\ref{back}). We implement the likelihood analysis using the geometrical data to put constraints on the cosmological and model parameters in PADE parameterisations. In Sect.(\ref{growth}), the growth of perturbations in PADE cosmologies is investigated. 
Then we perform an overall likelihood analysis including the geometrical + growth rate data to
place constraints and obtain the best fit values of the corresponding cosmological parameters. 
Finally we provide the main concussions in Sect.(\ref{conlusion}).

\section{PADE parameterisations}\label{sect:pade}
For an arbitrary function $f(x)$, the PADE approximate of order $(m,n)$ is given by the following rational function \citep{pade1892,baker96,Adachi:2011vu}
\begin{eqnarray}\label{padeO}
f(x)=\frac{a_0+a_1x+a_2x^2+...+a_nx^n}{b_0+b_1x+b_2x^2+...+b_nx^m}\;,
\end{eqnarray}
where exponents $(m,n)$ are positive and the coefficients $(a_{\rm i},b{\rm _i})$ are constants. 
Obviously, for $b_{\rm i}=0$ (with $i\ge 1$) 
the current approximation reduces to standard Taylor expansion. 
In this study we focus on three PADE parameterisations 
introduced as follows \citep[see also][]{Wei:2013jya}.

\subsection{PADE (I)}
Based on Eq. (\ref{padeO}), we first expand the EoS parameter $w_{\rm de}$ with respect to $(1-a)$ up to order $(1,1)$ 
as follows \citep[see also][]{Wei:2013jya}:
\begin{equation}\label{pade1}
w_{\rm de}(a)=\frac{w_0+w_{1}(1-a)}{1+w_{2}(1-a)}\;.
\end{equation}
From now on we will call the above formula  
as PADE (I) parameterisation. In terms of redshift $z$, Eq. (\ref{pade1}) is written as
\begin{equation}\label{pade1b}
w_{\rm de}(z)=\frac{w_0+(w_0+w_1)z}{1+(1+w_{2})z}\;.
\end{equation}
As expected for $w_2=0$ Eq. (\ref{pade1}) boils down to CPL parameterisation. Unlike CPL parameterisation, here the EoS 
parameter with $w_2\neq 0$
avoids the divergence at $a\to\infty$ (or equivalently at $z=-1$). 
Using Eq. (\ref{pade1}) we find the following special cases regarding the EoS parameter \citep[see also][]{Wei:2013jya}
\begin{figure*} 
	\centering
	\includegraphics[width=8cm]{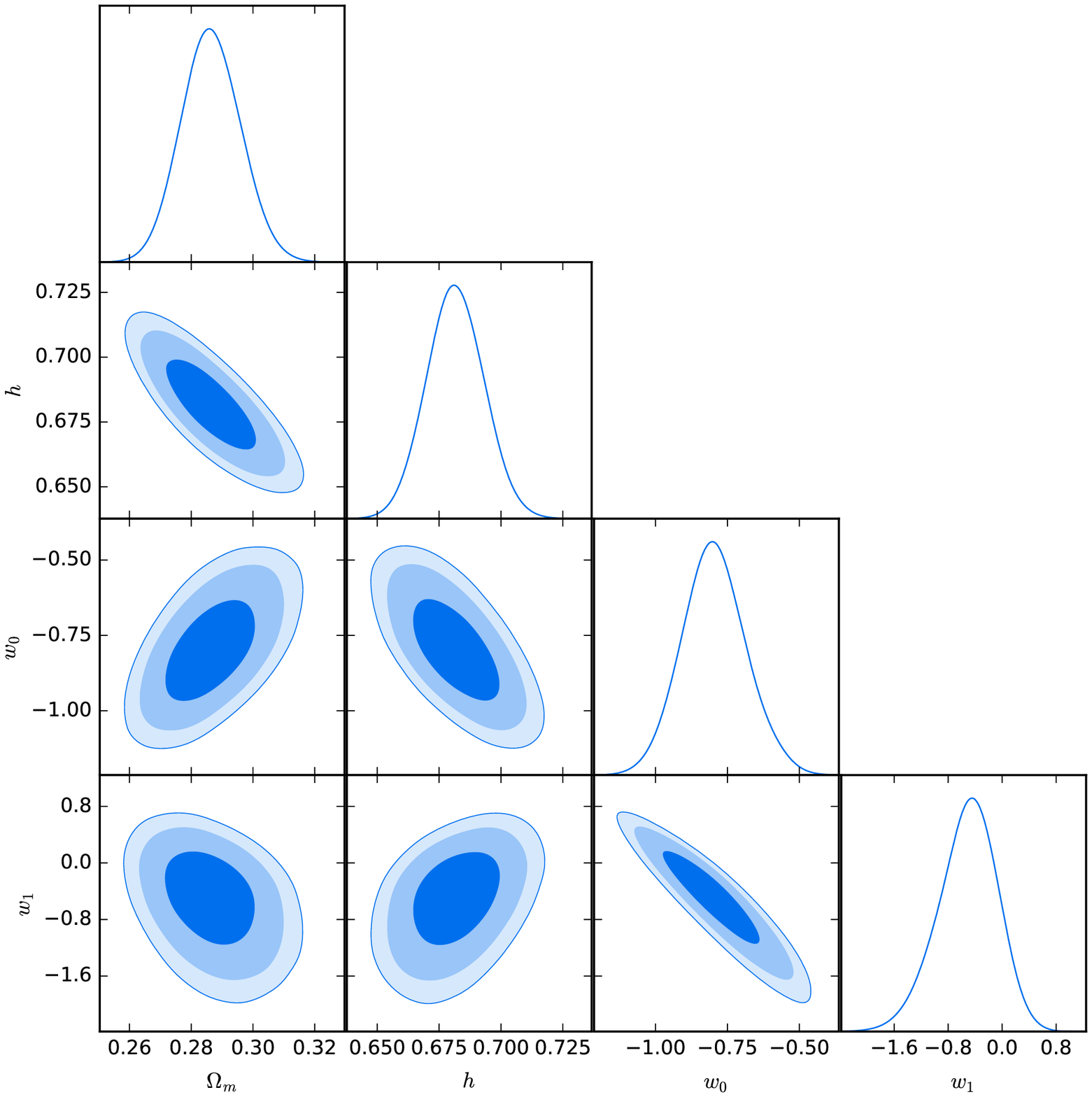}
	\includegraphics[width=8cm]{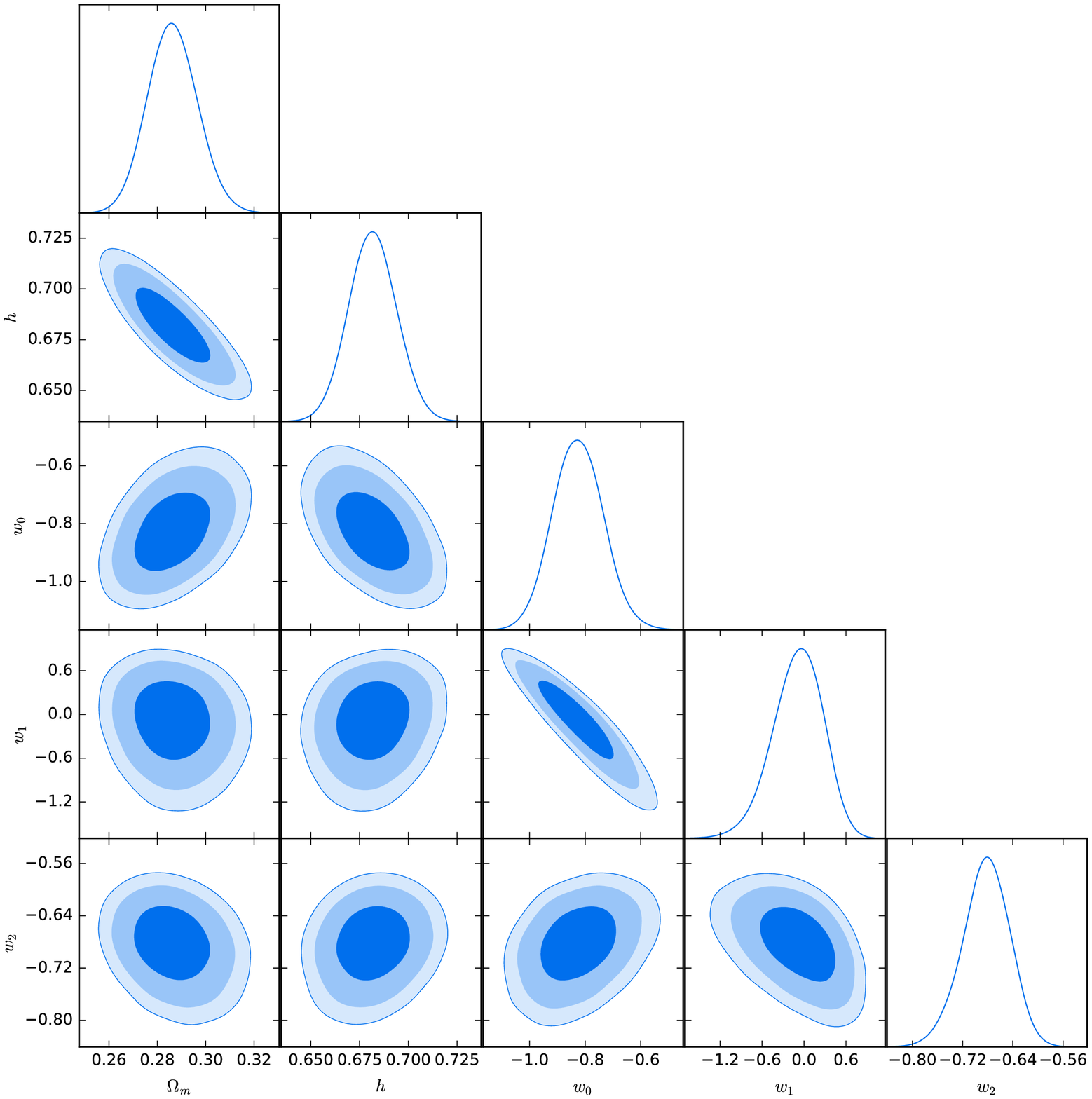}
	\includegraphics[width=8cm]{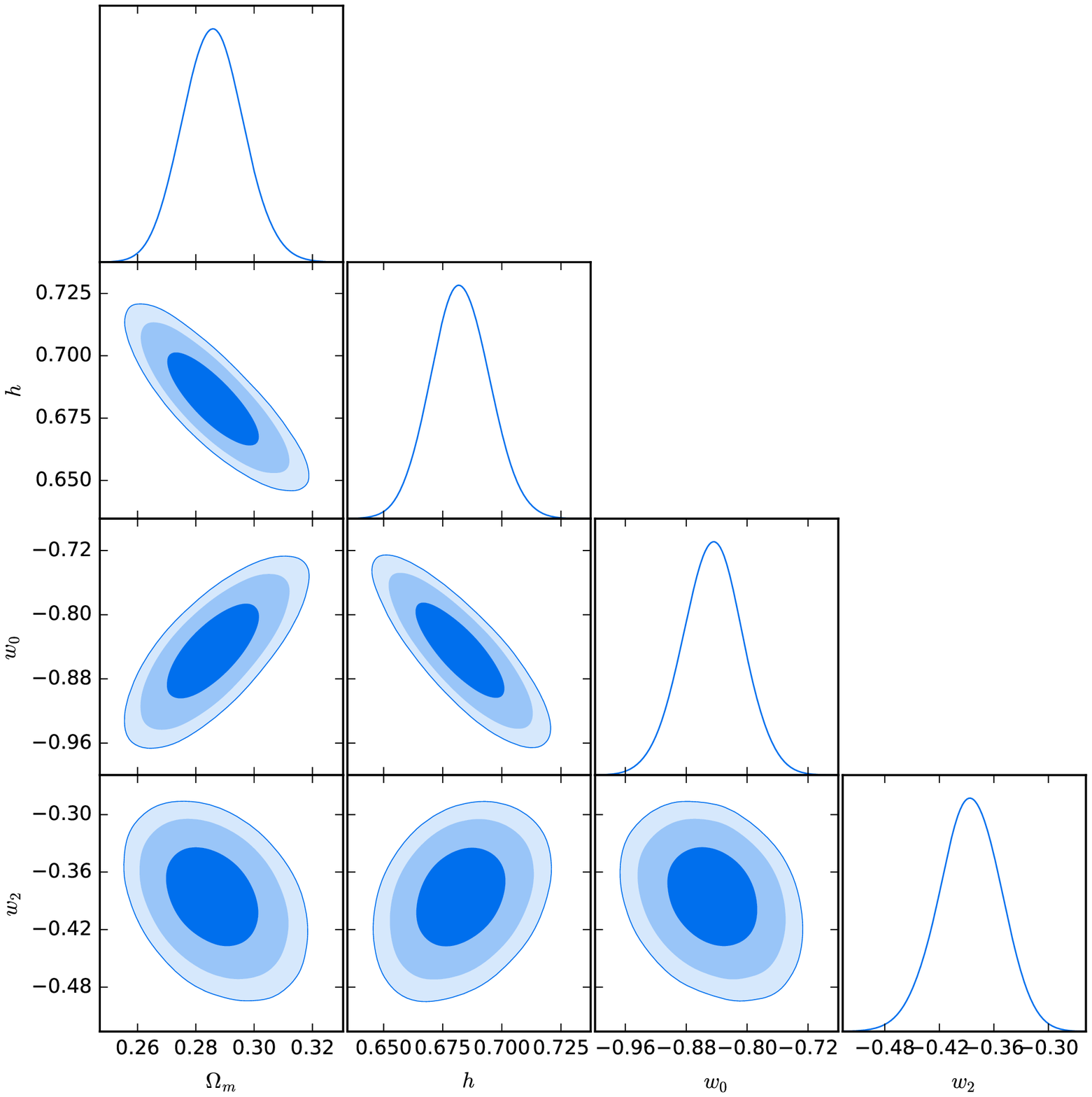}
	\includegraphics[width=8cm]{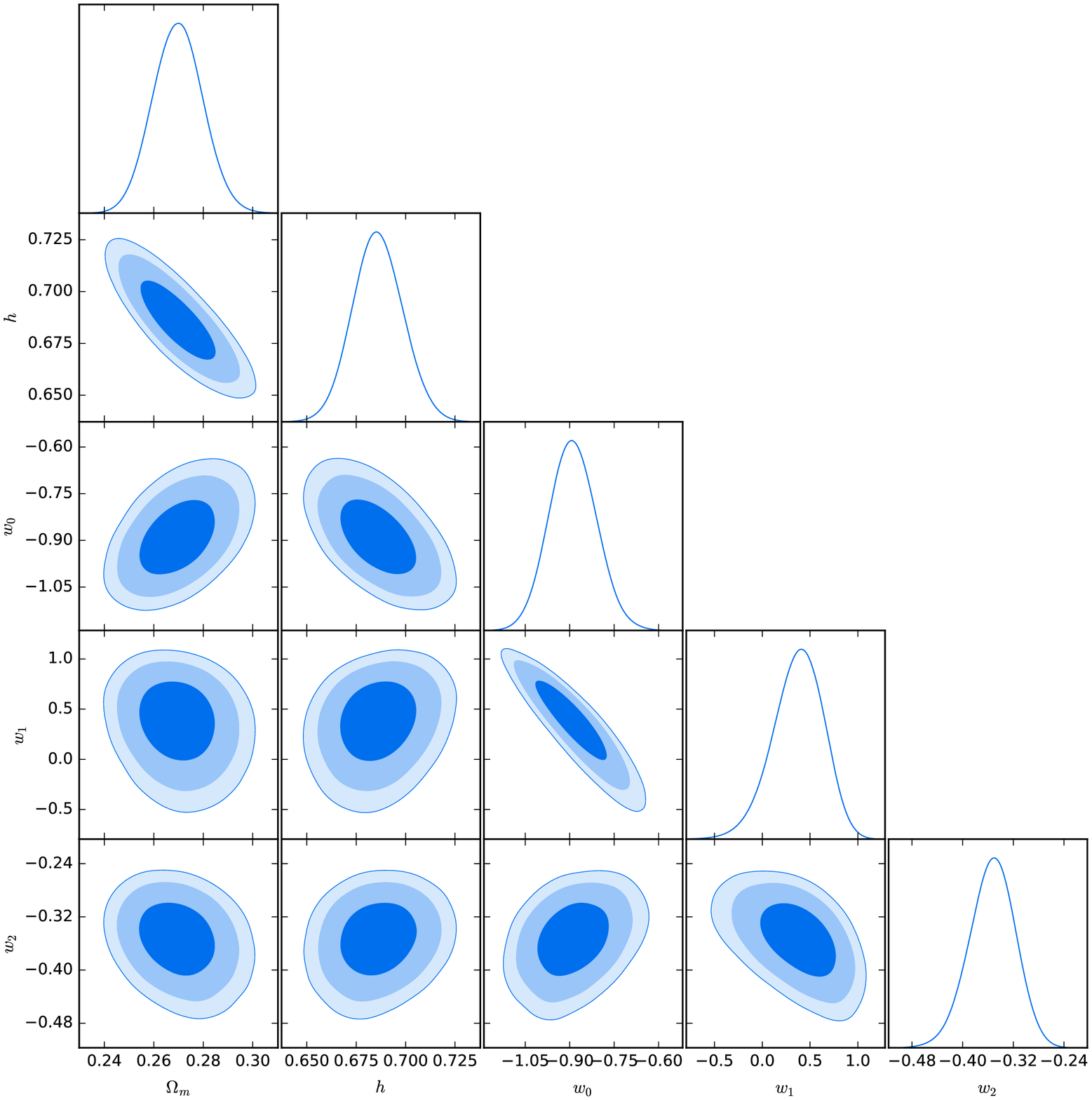}
	\caption{ The $1\sigma$, $2\sigma$ and $3\sigma$ likelihood contours 
for various cosmological parameters using the latest expansion data. The upper left (upper right) 
panel shows the results for CPL (PADE I) parameterisation. The lower left (lower right) panel shows the results 
for simplified PADE I (PADE II) parameterisation.}
	\label{fig:contour}
\end{figure*}
\begin{figure} 
	\centering
	\includegraphics[width=8cm]{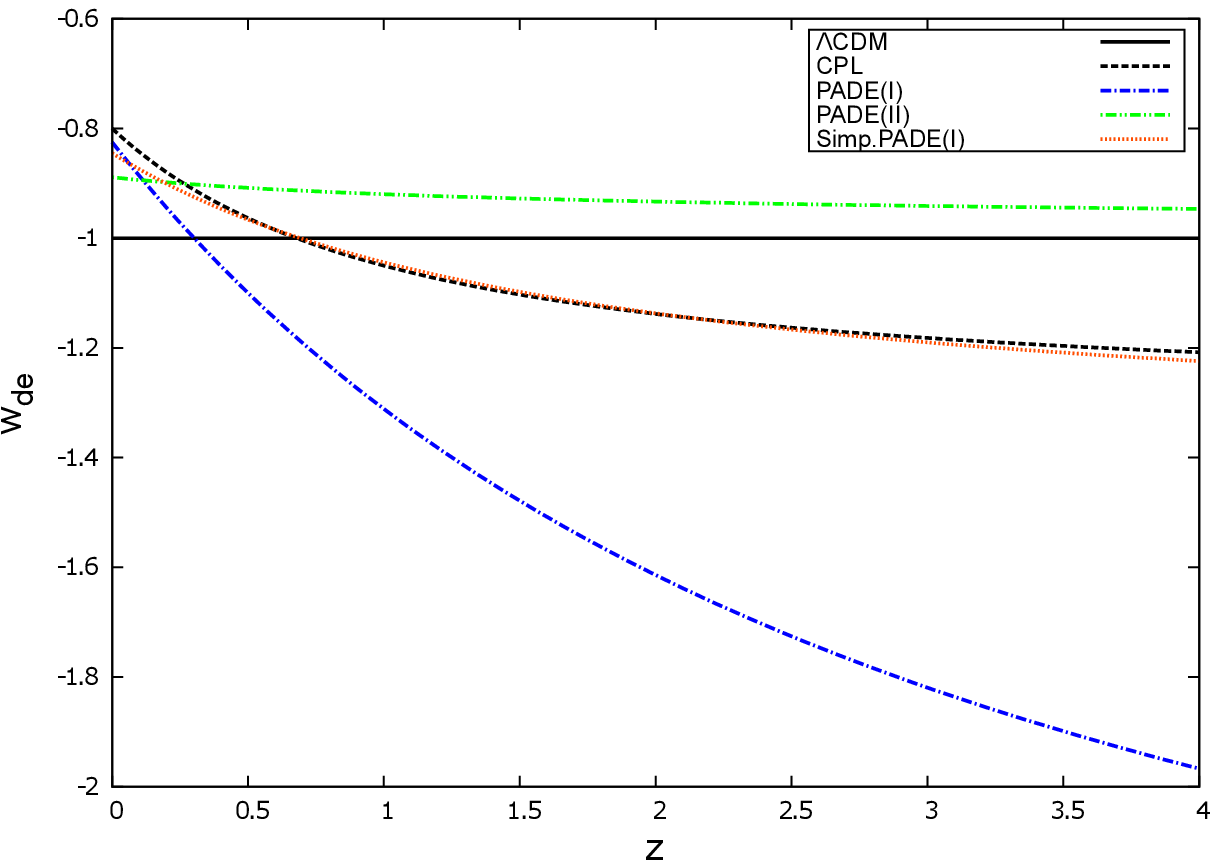}
	\includegraphics[width=8cm]{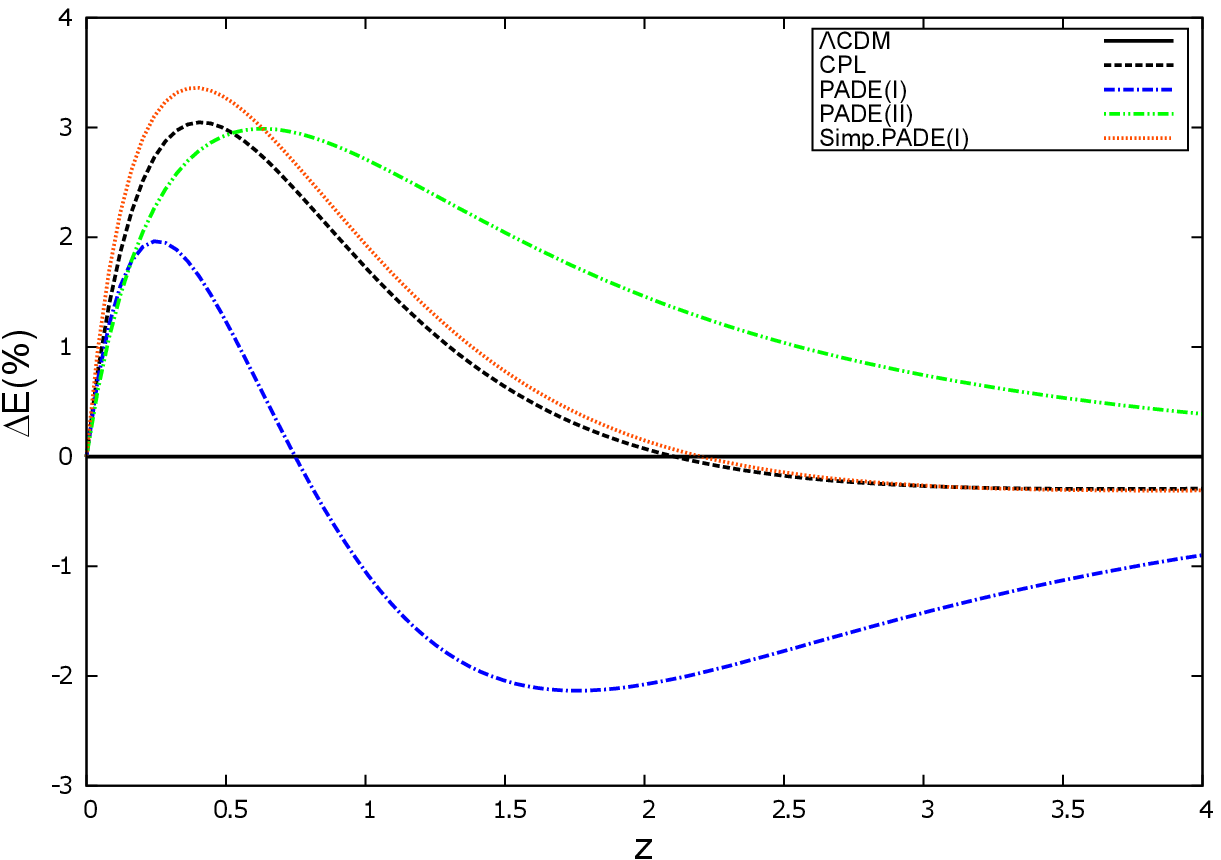}
	\includegraphics[width=8cm]{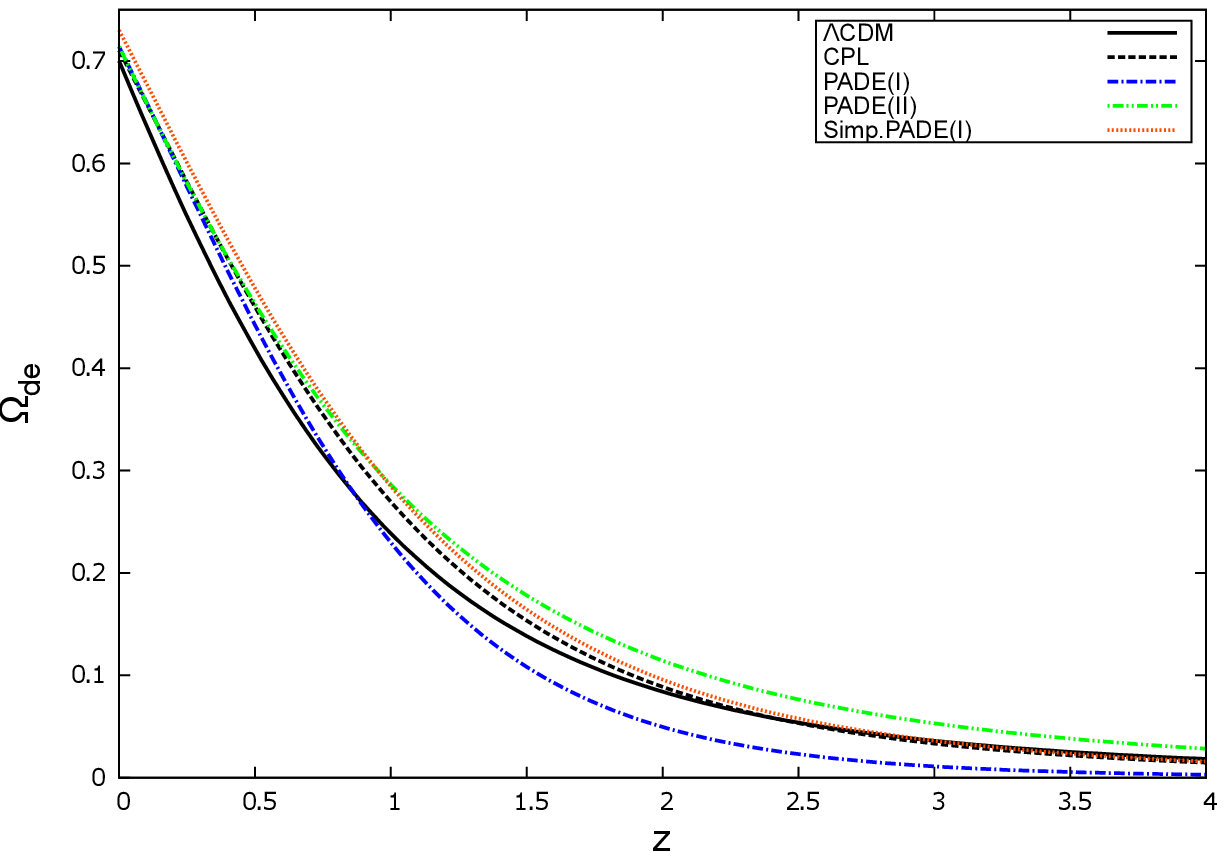}
	\caption{The redshift evolution of various cosmological quantities, namely dark energy EoS parameter $w_{\rm de}(z)$ ( top panel), relative deviation 
$\Delta E(\%)=[(E-E_{\rm \Lambda})/E_{\rm \Lambda}]\times 100$ (middle panel) and $\Omega_{\rm de}(z)$ ( bottom panel).
The different DE parameterisations are characterized by the colors and line-types presented
in the inner panels of the figure.}
	\label{fig:back}
\end{figure}

\begin{figure} 
	\centering
	\includegraphics[width=8cm]{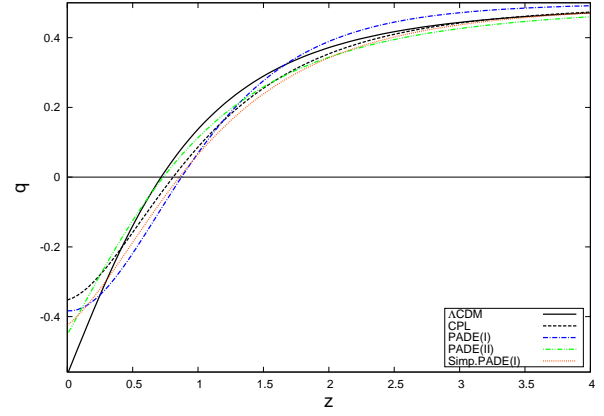}
	\caption{ The evolution of the 
deceleration parameter $q$ for different PADE parameterisations considered in this work. The CPL and $\Lambda$CDM models are shown for comparison. 
}
	\label{fig:q2}
\end{figure}

\begin{eqnarray}
w_{\rm de}=\left\{
\begin{array}{ll}
\frac{w_0+w_1}{1+w_2}\,,~~ & {\rm for}~~
a\to 0 ~(z\to\infty,~{\rm early~time})\,,\\[4mm]
w_0\,, & {\rm for}~~ a=1 ~(z=0,~{\rm present})\,,\\[4mm]
\frac{w_1}{w_2}\,, & {\rm for}~~ a\to\infty~
(z\to -1,~{\rm far~future})\,,
\end{array} \right.
\end{eqnarray}
where we need 
to set $w_2\not=0$ and $-1$. 
Therefore, we argue that PADE (I) formula is a well-behaved function in the 
range of $0\leq a\leq \infty$ (or equivalently at $-1\leq z\leq\infty$ ).\\

\subsection{simplified PADE (I)}
Clearly, PADE (I) approximation has three free parameters $w_0$, $w_1$ and $w_2$. 
Setting $w_1=0$ we provide a simplified version of PADE (I) parameterisation, namely
\begin{equation}\label{padesimp}
w_{\rm de}(a)=\frac{w_0}{1+w_{2}(1-a)}\;.
\end{equation}
Notice, that in order to avoid singularities in the cosmic expansion 
$w_{2}$ needs to lie in the interval $-1<w_2<0$.

\subsection{PADE (II)}
Unlike the previous cases, here 
the current parameterisation is written as a function of $N=\ln{a}$. 
In this context, the EoS parameter up to order ($1,1$) takes the form

\begin{equation}\label{pade2}
w_{\rm de}(a)=\frac{w_0+w_{1}\ln{a}}{1+w_{2}\ln{a}}\;,
\end{equation}
where $w_0$, $w_1$ and $w_2$ are constants \citep[see also][]{Wei:2013jya}. In PADE (II) parameterisation, we can easily show that
\begin{eqnarray}
w_{\rm de}=\left\{
\begin{array}{ll}
\frac{w_1}{w_2}\,,~~ & {\rm for}~~
a\to 0 ~(z\to\infty,~{\rm early~time})\,,\\[4mm]
w_0\,, & {\rm for}~~ a=1 ~(z=0,~{\rm present})\,,\\[4mm]
\frac{w_1}{w_2}\,, & {\rm for}~~ a\to\infty~
(z\to -1,~{\rm far~future})\,,
\end{array} \right.
\end{eqnarray}
Notice, that in order to avoid singularities at the above epochs we need to impose
$w_2\neq 0$.

\begin{table}
 \centering
 \caption{The statistical results for the various DE parameterisations used in the analysis. These results are based on the expansion data.
The concordance $\Lambda$CDM model is shown for comparison.}
\begin{tabular}{c  c  c c c c}
\hline \hline
 Model  & PADE I & simp. PADE I & PADE II& CPL&$\Lambda$CDM\\
 \hline
 $k$ & 6& 5 & 6 & 5 & 3\\
 \hline
$\chi^2_{\rm min}$ & 567.6 & 567.7 & 567.9 & 567.6 & 574.4\\
 \hline 
 AIC  & 579.6 & 577.7 & 579.9 & 577.6 & 580.4\\
 \hline
 BIC & 606.1 & 599.8 & 606.4 & 599.7 & 593.6\\
 & & & & & \\
 \hline \hline
\end{tabular}\label{tab:best}
\end{table}

\section{background history in PADE parameterisations}
\label{back}
In this section based on the aforementioned parameterisations we study
the background evolution in PADE cosmologies. 
Generally speaking, for isotropic and homogeneous spatially flat FRW cosmologies, driven by radiation, 
non-relativistic matter and an exotic fluid with an equation of state
$p_{\rm de}=w_{\rm de}\rho_{\rm de}$, the first Friedmann equation reads 
\begin{eqnarray}\label{frid1}
H^2=\frac{8\pi G}{3}(\rho_{\rm r}+\rho_{\rm m}+\rho_{\rm de})\;,
\end{eqnarray}
where $H\equiv {\dot a}/a$ is the Hubble parameter, $\rho_{\rm r}$, $\rho_{\rm m}$ and $\rho_{\rm de}$ are 
the energy densities of radiation, dark matter and DE, respectively.
In the absence of interactions among the three fluids the corresponding 
energy densities satisfy the following differential equations
 \begin{eqnarray}\label{continuity}
 && \dot{\rho_{\rm r}}+4H\rho_{\rm r}=0\;,\label{radiation}\\
&&\dot{\rho_{\rm m}}+3H\rho_{\rm m}=0\;,\label{matter}\\
&&\dot{\rho_{\rm de}}+3H(1+w_{\rm de})\rho_{\rm de}=0\;\label{de},
 \end{eqnarray}
where the over-dot denotes a derivative with respect to cosmic time $t$. 
Based on Eqs. (\ref{radiation}) and (\ref{matter}), it is easy to derive the evolution of radiation and pressure-less matter, 
namely $\rho_{\rm r}=\rho_{\rm r0}a^{-4}$ and $\rho_{\rm m}=\rho_{\rm m0}a^{-3}$. 
Inserting Eqs . (\ref{pade1}), (\ref{padesimp}) and (\ref{pade2}) into equation (\ref{de}), we can obtain 
the DE density of the current PADE parameterisations 
\citep[see also][]{Wei:2013jya}

\begin{eqnarray}
\rho_{\rm de}^{(\rm PADE I)}=\rho^{(0)}_{\rm de} a^{-3(\frac{1+w_0 + w_1+ w_2}{1+w_2})} [1+w_2 (1-a)]^{-3(\frac{w_1- w_0 w_2}{w_2 (1+w_2)})}\;,\label{rho-pade1}\\
\rho_{\rm de}^{(\rm simp. PADE I)}=\rho^{(0)}_{\rm de} a^{-3(\frac{1+w_0+ w_2}{1+w_2})} [1+w_2 (1-a)]^{-3(\frac{- w_0 w_2}{w_2 (1+w_2)})}\;,\label{rho-simpade1}\\
\rho_{\rm de}^{(\rm pade II)}=\rho^{0}_{\rm de} a^{-3(\frac{w_1+ w_2}{w_2})} (1+w_2 \ln a)^{3(\frac{w_1- w_0 w_2}{{w_2}^2})}\;.\label{rho-pade2}
\end{eqnarray}
Also, combining Eqs.(\ref{rho-pade1}, \ref{rho-simpade1}, \ref{rho-pade2}) and Eq.(\ref{frid1}) we derive the dimensionless Hubble 
parameter $E=H/H_0$ \citep[see also][]{Wei:2013jya}. Specifically, we find 

\begin{eqnarray}
&&E^{2}_{\rm PADE I}=\Omega_ {\rm r0} a^{-4}+\Omega_{\rm m0} a^{-3} + (1-[\Omega_{\rm r0}+\Omega_{\rm m0}]) \times \nonumber \\
&&a^{-3(\frac{1+w_0+ w_1 + w_2}{1+w_2})} \times (1+w_2 - a w_2)^{-3(\frac{w_1 -w_0 w_2}{w_2(1+w_2)})} \;,\label{Epade1}\\
&&E^{2}_{\rm simPADE I}=\Omega_{\rm r0} a^{-4}+\Omega_{\rm m0} a^{-3} + (1-[\Omega_{\rm r0}+\Omega_{\rm m0}]) \times \nonumber \\ &&a^{-3(\frac{1+w_0+ w_2}{1+w_2})} \times(1+w_2 - a w_2)^{-3(\frac{ -w_0 w_2}{w_2(1+w_2)})} \;,\label{Esimpade1}\\
&&E^{2}_{\rm PADE II}=\Omega_{\rm r0} a^{-4}+\Omega_{\rm m0} a^{-3} +(1-[\Omega_{\rm r0}+\Omega_{\rm m0}]) \times \nonumber\\ &&a^{-3(\frac{w_1 + w_2}{w_2})}\times (1+w_2 \ln a)^{3(\frac{w_1 -w_0 w_2}{{w_2}^2})} \;,\label{Epade2}
\end{eqnarray}
where $\Omega_{\rm m0}$ (density parameter), 
$\Omega_{\rm r0}$ (radiation parameter) and 
$\Omega_{\rm de0}=1-\Omega_{\rm m0}-\Omega_{\rm r0}$ (dark energy parameter).
Moreover, following the above lines in the case of CPL parameterisation we have 
\begin{eqnarray}\label{rho-cpl}
\rho^{\rm CPL}_{\rm de}=\rho^{(0)}_{\rm de}a^{-3(1+w_0+w_1)}\exp\{-3w_1(1-a)\}
\end{eqnarray}
and 
\begin{eqnarray}
&&E^{2}_{\rm CPL}=\Omega_{\rm r0} a^{-4}+\Omega_{\rm m0} a^{-3} + (1-\Omega_{\rm m0}-\Omega_{\rm r0}) \times \nonumber \\
&&a^{-3(1+w_0+ w_1)}\exp[{-3 w_1 (1-a)}]\;,\label{Ecpl}
\end{eqnarray}
Bellow, we study the performance of PADE 
cosmological parameterisation against the
latest observational data. 
Specifically, we implement a statistical analysis using the background expansion data 
including those of SnIa \citep{Union2.1:2012}, BAO \citep{Beutler:2011hx,Padmanabhan:2012hf,
Anderson:2012sa,Blake:2011en}, CMB \citep{Hinshaw:2012aka}, BBN \citep{Serra:2009yp,Burles:2000zk}, Hubble data \citep{Moresco:2012jh,Gaztanaga:2008xz,Blake:2012pj,Anderson:2013zyy}. For more details concerning the expansion data, the $\chi^2(\textbf{p})$ function, the Markov 
chain Monte Carlo (MCMC) analysis, the Akaike information criterion (AIC) and the Bayesian information 
criterion (BIC) we refer the reader to \cite{Mehrabi:2015hva} \citep[see also][]{Basilakos:2009wi,Hinshaw:2012aka,Mehrabi:2015kta,Mehrabi:2016exz,Malekjani:2016edh}. 
In this framework, the joint likelihood function is the product 
of the individual likelihoods:
\begin{equation}\label{eq:like-tot}
 {\cal L}_{\rm tot}({\bf p})={\cal L}_{\rm sn} \times {\cal L}_{\rm bao} \times {\cal L}_{\rm cmb} \times {\cal L}_{\rm h} \times
 {\cal L}_{\rm bbn}\;,
\end{equation}
which implies that the total chi-square $\chi^2_{\rm tot}$ is given by:
\begin{equation}\label{eq:like-tot_chi}
 \chi^2_{\rm tot}({\bf p})=\chi^2_{\rm sn}+\chi^2_{\rm bao}+\chi^2_{\rm cmb}+\chi^2_{\rm h}+\chi^2_{\rm bbn}\;,
\end{equation}
where the statistical vector ${\bf p}$ includes the free parameters of the model. 
In our work the above vector becomes 
(a) ${\bf p}=\{\Omega_{\rm DM0},\Omega_{\rm b0}, h,w_0,w_1,w_2\}$ for PADE (I) and (II) parameterisations, (b) 
${\bf p}=\{\Omega_{\rm DM0},\Omega_{\rm b0}, h,w_0,w_2\}$ for simplified PADE (I) 
and (c) ${\bf p}=\{\Omega_{\rm DM0},\Omega_{\rm b0}, h,w_0,w_1\}$ in the case of CPL parameterisation. 
Notice that we utilize $\Omega_{\rm m0}=\Omega_{\rm DM0}+\Omega_{\rm b0}$ and $h=H_{0}/100$, while 
the energy density of radiation is fixed to $\Omega_{\rm r0}=2.469\times 10^{-5}h^{-2}(1.6903)$ \citep{Hinshaw:2012aka}. 

Additionally, we utilize the well know information criteria, namely 
AIC \citep{Akaike:1974} and BIC \citep{Schwarz:1974} in
order to test the statistical performance of the cosmological models themselves.
In particular, AIC and BIC are given by 
\begin{eqnarray}
{\rm AIC} = -2 \ln {\cal L}_{\rm max}+2k\;,\nonumber\\
{\rm BIC} = -2 \ln {\cal L}_{\rm max}+k\ln N\;,
\end{eqnarray}
where $k$ is the number of free parameters and $N$ 
is the total number of observational data points. The results of our statistical analysis 
are presented in Tables (\ref{tab:best}) and (\ref{tab:bestfit}) respectively. 
Although the current DE parameterisations provide low AIC values with respect to those
of $\Lambda$CDM, we find $\Delta {\rm AIC}={\rm AIC}-{\rm AIC}_{\rm \Lambda}<4$ hence, 
the DE parameterisations explored in this study are consistent with the expansion data.
In order to visualize the solution space of the model parameters
in Fig.(\ref{fig:contour}) we present 
the $1\sigma$, $2\sigma$ and $3\sigma$ confidence levels for 
various parameter pairs.
Using the best fit model parameters [see Table \ref{tab:bestfit}] 
in Fig. (\ref{fig:back}) we plot
the redshift evolution of $w_{\rm de}$ (upper panel), $\Delta E(\%)=[(E-E_{\rm \Lambda})/E_{\rm \Lambda}]\times 100$ (middle panel) 
and $\Omega_{\rm de}$ (lower panel). 
The different parameterisations are characterized by the colors and line-types presented
in the caption of Fig. (\ref{fig:back}).  
We find that the EoS parameter of PADE II evolves only in the
quintessence regime ($-1<w_{\rm de}<-1/3$). 
For the other DE parameterisations we observe that 
$w_{\rm de}$ 
varies in the phantom region ($w_{\rm de}<-1$) at high redshifts, while 
it enters in the quintessence regime ($-1<w_{\rm de}<-1/3$) at relatively low redshifts. Notice, that the present value of $w_{\rm de}$ 
can be found in Table (\ref{tab:bestfit}).
From the middle panel of Fig. (\ref{fig:back}), we observe that the relative difference
$\Delta E$ is close to $2-3.5\%$ at low redshifts ($z \sim 0.5$), 
while in the case of PADE (II)  we always have $E_{\rm PADE II}(z)>E_{\rm \Lambda}(z)$.
Lastly, in the bottom panel of Fig.(\ref{fig:back}) we show the evolution of $\Omega_{\rm de}$, where its current value
can be found in Table (\ref{tab:bestfit}).
As expected, $\Omega_{\rm de}$ tends to zero at high redshifts since matter dominates 
the cosmic fluid. In the case of PADE 
parameterisations we observe that $\Omega_{\rm de}$ is 
larger than that of the usual $\Lambda$ cosmology.

Finally, we would like to estimate the transition redshift 
 $z_{\rm tr}$ of the PADE parameterisations by utilizing the deceleration parameter
$q(z)=-1-\dot{H}/H^2$. Following, standard lines it is easy to show  
\begin{eqnarray}
	\frac{\dot{H}}{H^2}=-\frac{3}{2}\Big(1+w_{\rm de}(z)\Omega_{\rm de}(z)\Big)
	\end{eqnarray}
which implies that
\begin{eqnarray}\label{eq:q2}
		q(z)=\frac{1}{2}+\frac{3}{2}w_{\rm de}(z)\Omega_{\rm de}(z)
\end{eqnarray}
		Using the best fit values of Table (\ref{tab:bestfit}),  
we plot in Fig. (\ref{fig:q2}) the evolution of $q$ for the current DE parameterisations.
In all cases, including that of $\Lambda$CDM, $q$ tends to 1/2 at early enough times. 
This is expected since the universe is matter dominated 
($\Omega_{\rm de} \simeq 0$) at high redshifts. 
Now solving the $q(z_{\rm tr})=0$ we can derive the transition redshift, namely the epoch at which 
the expansion of the universe starts to accelerate.
In particular, we find 
$z_{\rm tr}=0.86$ (PADE I), $z_{\rm tr}=0.84$ (simplified PADE ), 
$z_{\rm tr}=0.72$ (PADE II), $z_{\rm tr}=0.80$ (CPL) and $z_{\rm tr}=0.71$ ($\Lambda$CDM). 
The latter results are in good agreement with the measured $z_{\rm tr}$ based on 
the cosmic chronometer $H(z)$ data \cite{Farooq:2016zwm} \citep[see also][]{Capozziello:2014zda,Capozziello:2015rda}.

\begin{table*}
	\centering
	
	\caption{A summary of the best-fit parameters for the various DE parameterisations using the background data.
	}
	\begin{tabular}{c  c  c c c c}
		\hline \hline
		Model  & PADE I & simplified PADE I & PADE II& CPL&$\Lambda$CDM\\
		\hline 
		$\Omega_{\rm m}^{(0)}$ & $0.286\pm 0.010$ & $0.270\pm 0.010$ & $0.2864\pm 0.0096$ & $0.2896\pm 0.0090$ & $0.2891\pm 0.0090$\\
		\hline
		$ h $& $0.682\pm 0.012$&$0.682\pm 0.012$ &$0.686\pm 0.013$&$0.682\pm 0.012$&$0.6837\pm 0.0084$\\
		\hline
		$ w_0 $ & $-0.825\pm 0.091$&$-0.845\pm 0.039$ & $-0.889\pm 0.080$& $-0.80\pm 0.11$& $-$\\
		\hline
		$ w_1 $ & $-0.09^{+0.39}_{-0.32}$&$-$ & $0.37^{+0.29}_{-0.23}$& $-0.51^{+0.48}_{-0.38}$& $-$\\
		\hline
		$ w_2 $ & $-0.683^{+0.040}_{-0.034}$&$-0.387\pm 0.034$ & $-0.353^{+0.038}_{-0.034}$& $-$& $-$\\
		\hline
		$ w_{\rm de}(z=0) $ & $-0.825$&$-0.845$ & $-0.889$& $-0.80$& $-1.0$\\
		\hline
		$ \Omega_{\rm de}(z=0) $ & $0.714$ & $0.730$ & $0.7136$ & $0.7104$ & $0.7109$\\
		\hline \hline
	\end{tabular}\label{tab:bestfit}
\end{table*}

\begin{figure*} 
	\centering
	\includegraphics[width=8cm]{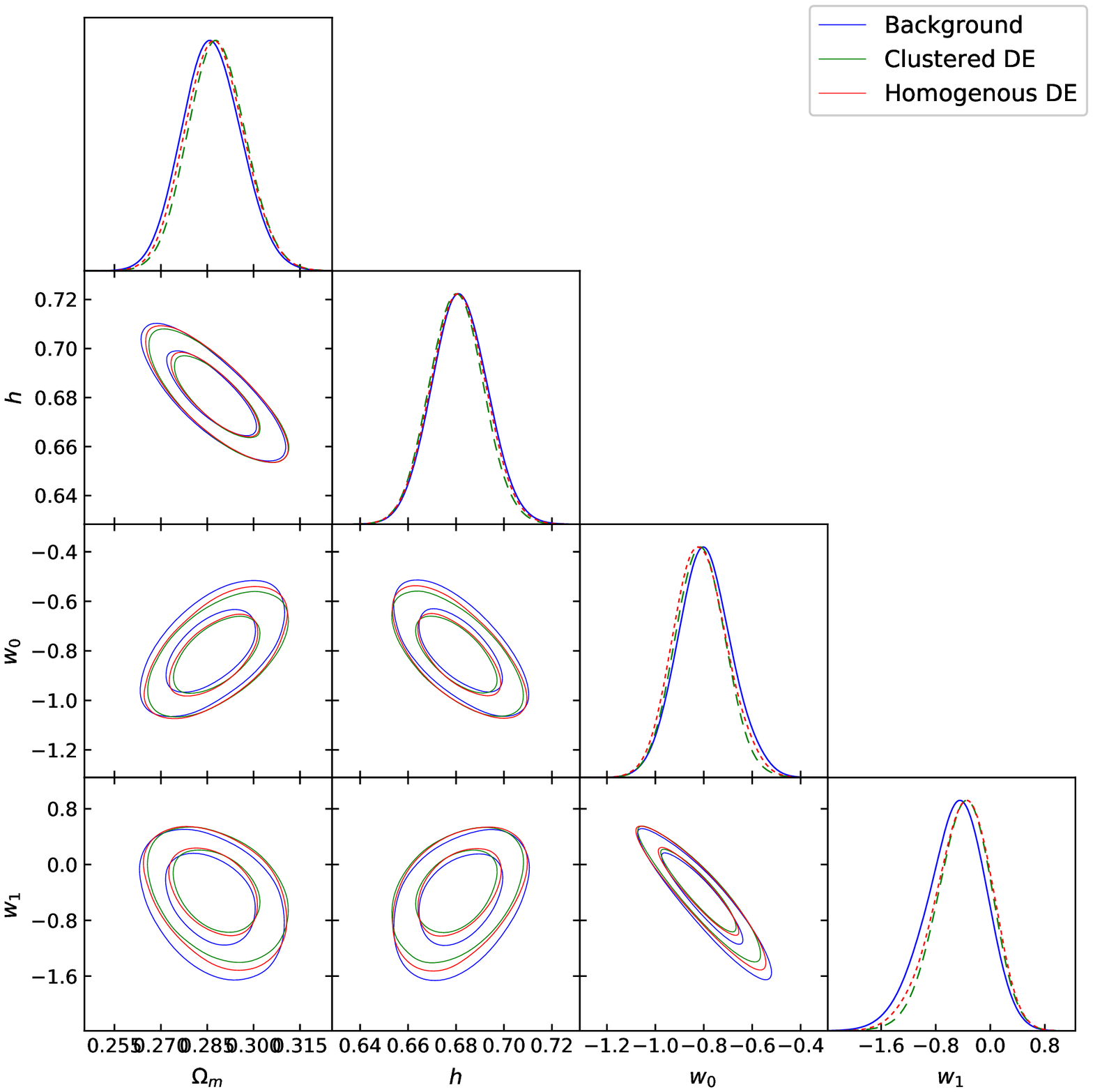}
	\includegraphics[width=8cm]{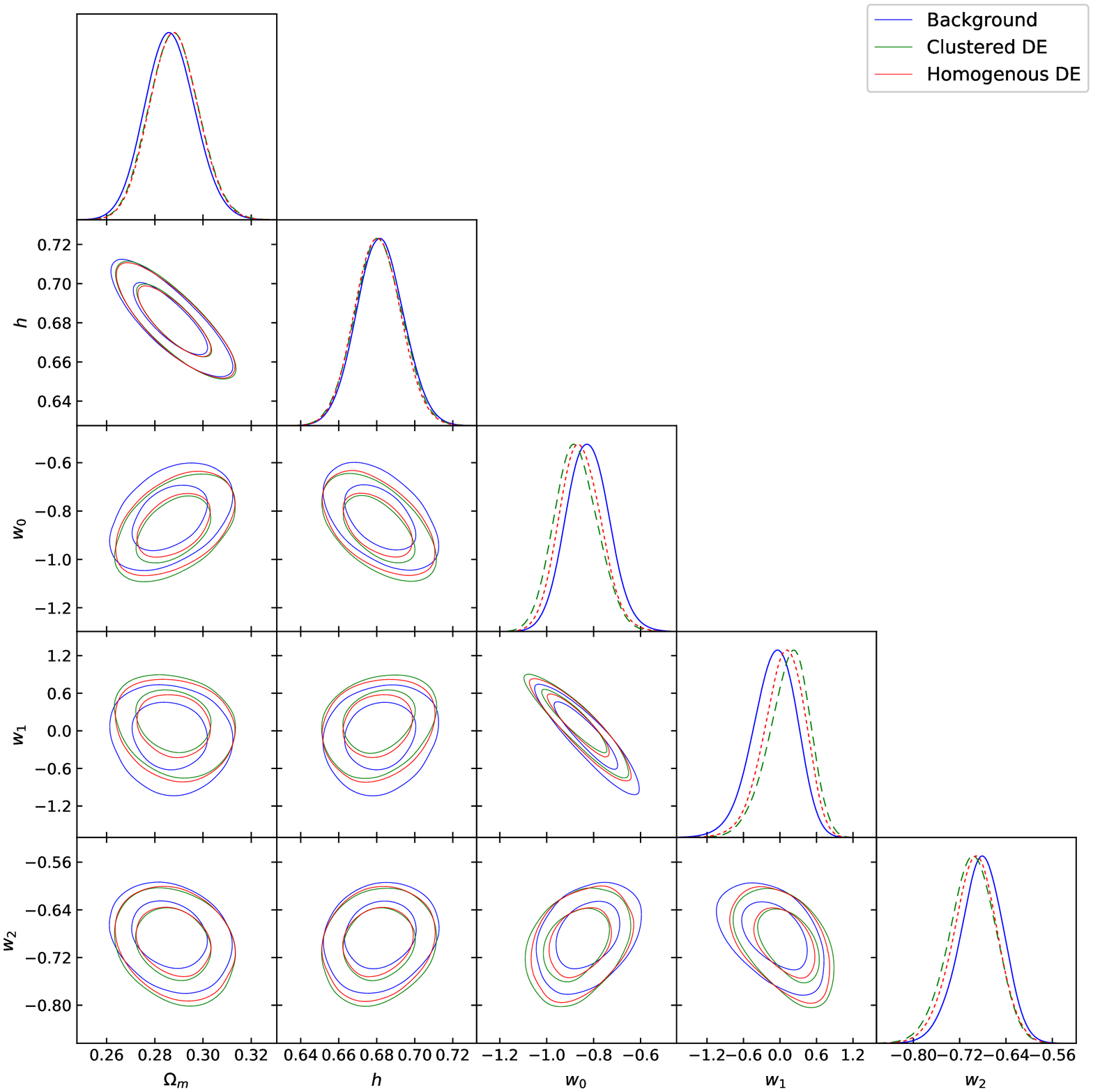}
	\includegraphics[width=8cm]{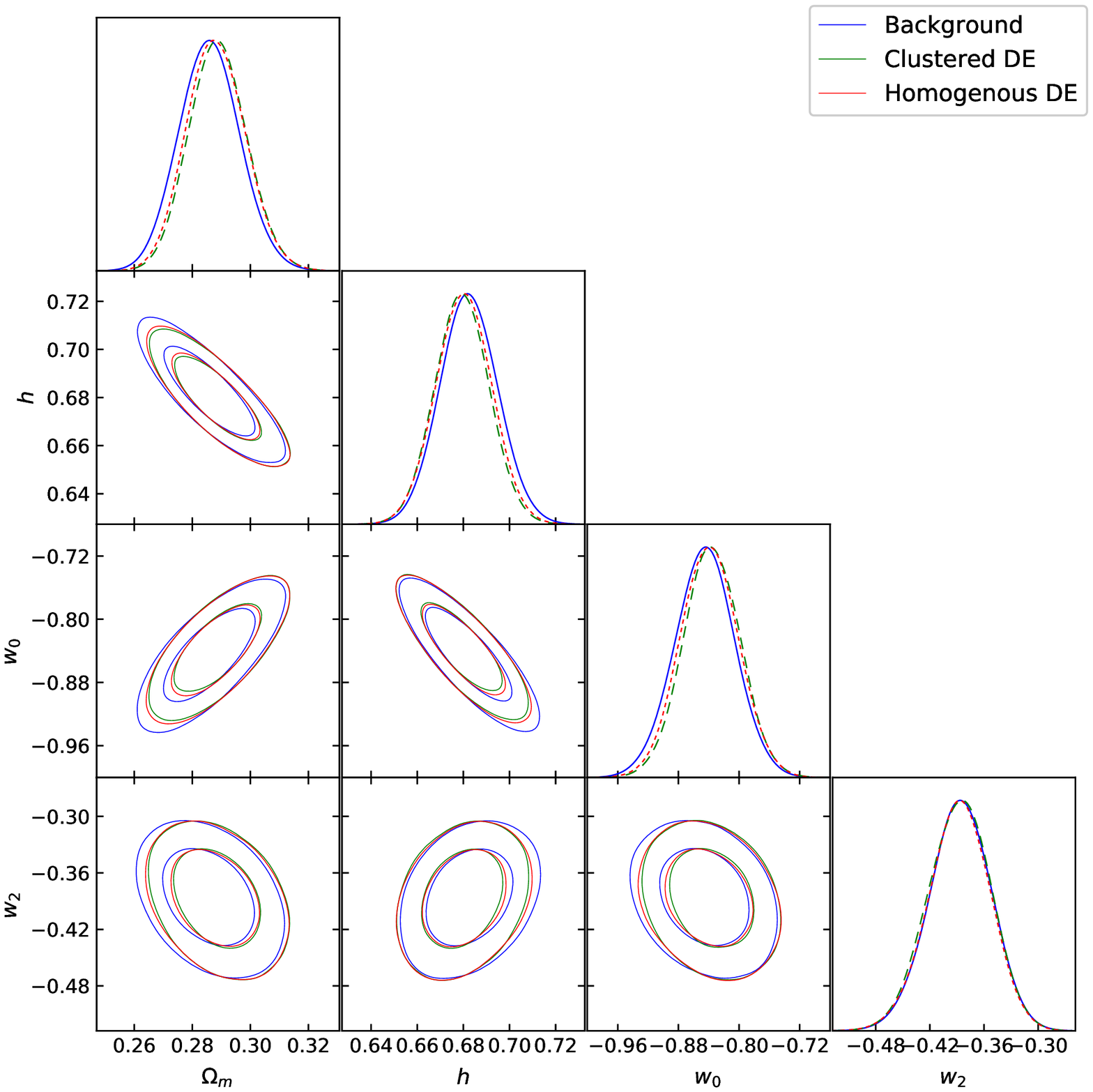}
	\includegraphics[width=8cm]{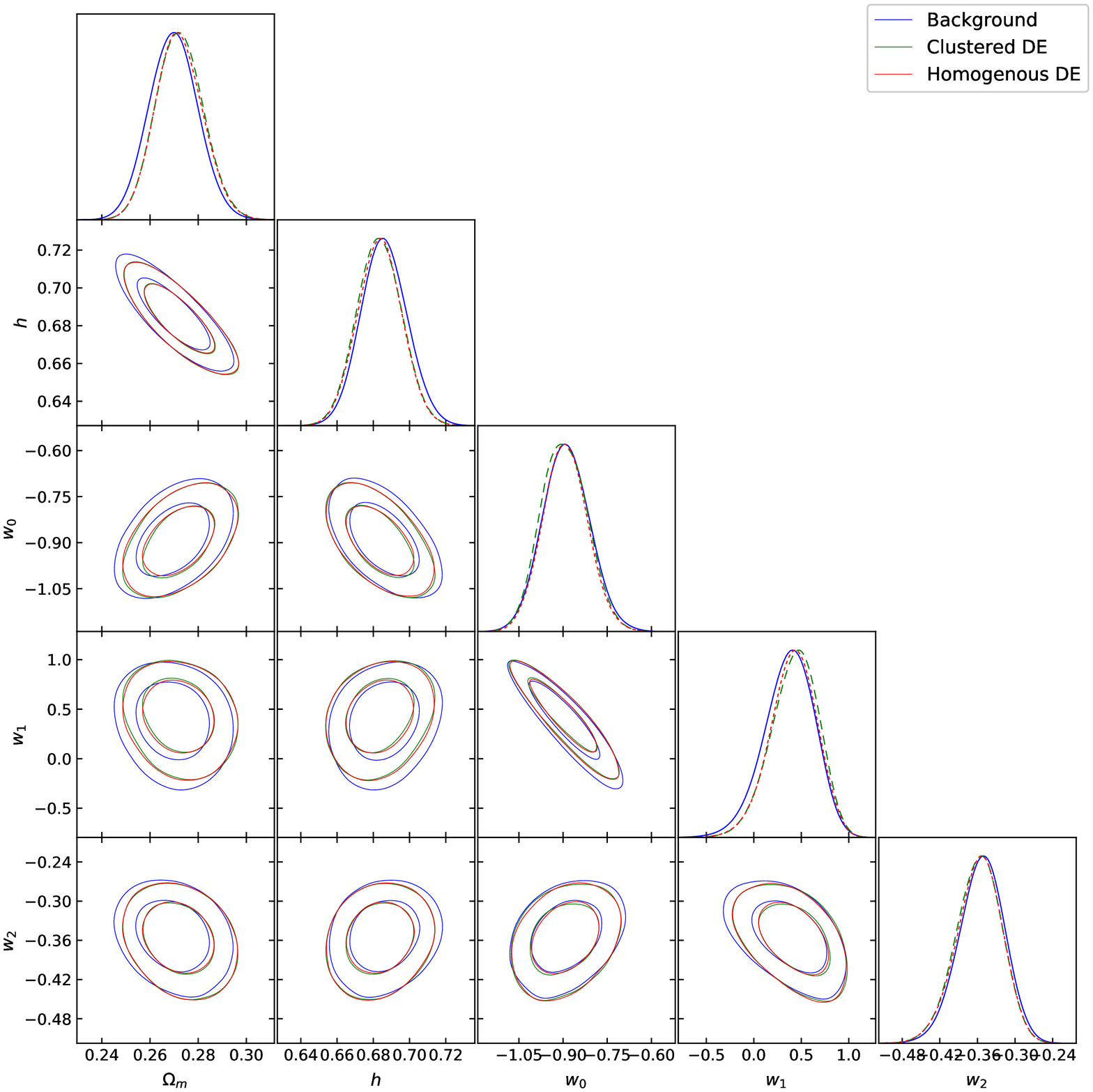}
	\caption{ The $1\sigma$ and $2\sigma$ likelihood contours 
for various planes using the solely expansion data (blue), combined expansion and growth rate data for clustered (green) and homogeneous (red) DE parameterisations. The upper left (upper right) shows the results for CPL (PADE I) parameterisation. The lower left (lower right) shows the results for simplified PADE I (PADE II) parameterisation.}
	\label{fig:contour_all}
\end{figure*}

\begin{table*}
	\centering
	\caption{The statistical results for homogeneous (clustered) DE parameterisations used in the analysis. These results are based on 
the expansion+growth rate data. The concordance $\Lambda$CDM model is shown for comparison.}
	\begin{tabular}{c  c  c c c c}
		\hline \hline
		Model  & PADE I & simplified PADE I & PADE II& CPL&$\Lambda$CDM\\
		\hline
		$k$ & 7& 6 & 7 & 6 & 4\\
		\hline
		$\chi^2_{\rm min}$ & 576.4(576.5) & 576.4(576.7) & 576.9(577.1) & 576.5(576.7) &582.6 \\
		\hline 
		AIC  & 590.4(590.5) & 588.4(588.7) & 590.9(591.1) & 588.5(588.7) &590.6 \\
		\hline
		BIC & 621.6(621.7) & 615.1(615.4) & 622.1(622.3) & 615.2(615.4) &608.4 \\
		& & & & & \\
		\hline \hline
	\end{tabular}\label{tab:best2}
\end{table*}

\begin{table*}
	\centering
	\caption{A summary of the best-fit parameters for homogeneous (clustered) DE parameterisations using the background+growth rate data.
}
	\resizebox{\textwidth}{!}{%
		\begin{tabular}{c  c  c c c c}
			\hline \hline
			Model  & PADE I & simplified PADE I & PADE II& CPL&$\Lambda$CDM\\
			\hline 
			$\Omega_{\rm m}^{(0)}$ & $0.288\pm 0.010~(0.288\pm 0.010)$ & $0.288\pm 0.010~(0.2888\pm 0.0099)$&$0.2721\pm 0.0097~(0.2723\pm 0.0098)$& $0.2875\pm 0.0095~(0.2882\pm 0.0093)$ & $0.2902\pm 0.0090$\\
			\hline
			$ h $& $0.681\pm 0.012~(0.681\pm 0.012)$&$0.680\pm 0.012~(0.679\pm 0.012)$&$0.684\pm 0.012~(0.683\pm 0.012)$&$0.681\pm 0.011~(0.680\pm 0.011)$&$0.6833\pm 0.0084$\\
			\hline
			$ w_0 $ & $-0.856\pm 0.088~(-0.874^{+0.086}_{-0.097})$&$-0.839\pm 0.038~(-0.836\pm 0.037)$ & $-0.893\pm 0.075~(-0.896\pm 0.078)$& $-0.81^{+0.10}_{-0.12}~(-0.81\pm 0.10)$& $-$\\
			\hline
			$ w_1 $ & $0.07^{+0.37}_{-0.29}~(0.14^{+0.38}_{-0.29})$&$-$ & $0.41^{+0.26}_{-0.22}~(0.43^{+0.27}_{-0.22})$& $-0.41^{+0.46}_{-0.37}~(-0.39^{+0.42}_{-0.36})$& $-$\\
			\hline
			$ w_2 $ & $-0.694^{+0.040}_{-0.036}~(-0.699^{+0.042}_{-0.038})$&$-0.388\pm 0.034~(-0.388\pm 0.035)$ & $-0.357^{+0.039}_{-0.033}~(-0.358^{+0.038}_{-0.034})$& $-$& $-$\\
			\hline
			$ \sigma_8 $ & $0.751\pm 0.015~(0.755\pm 0.016)$&$0.751\pm 0.015~(0.758\pm 0.015)$& $0.771\pm 0.015~(0.771\pm 0.016)$& $0.751\pm 0.015~(0.756\pm 0.015)$& $0.744\pm 0.014$\\
			\hline
			$ w_{\rm de}(z=0) $ &$-0.856(-0.874)$&$-0.839(-0.836)$ & $-0.893(-0.896)$& $-0.81(-0.81)$& $-1.0$\\
			\hline
			$ \Omega_{\rm de}(z=0) $ & $0.712(0.712)$ & $0.712(0.7112)$& $0.7279(0.7277)$ & $0.7125(0.7118)$ & $0.7098$\\
			\hline \hline
		\end{tabular}\label{tab:bestfit2}
	}
\end{table*}
\begin{figure} 
	\centering
	\includegraphics[width=8cm]{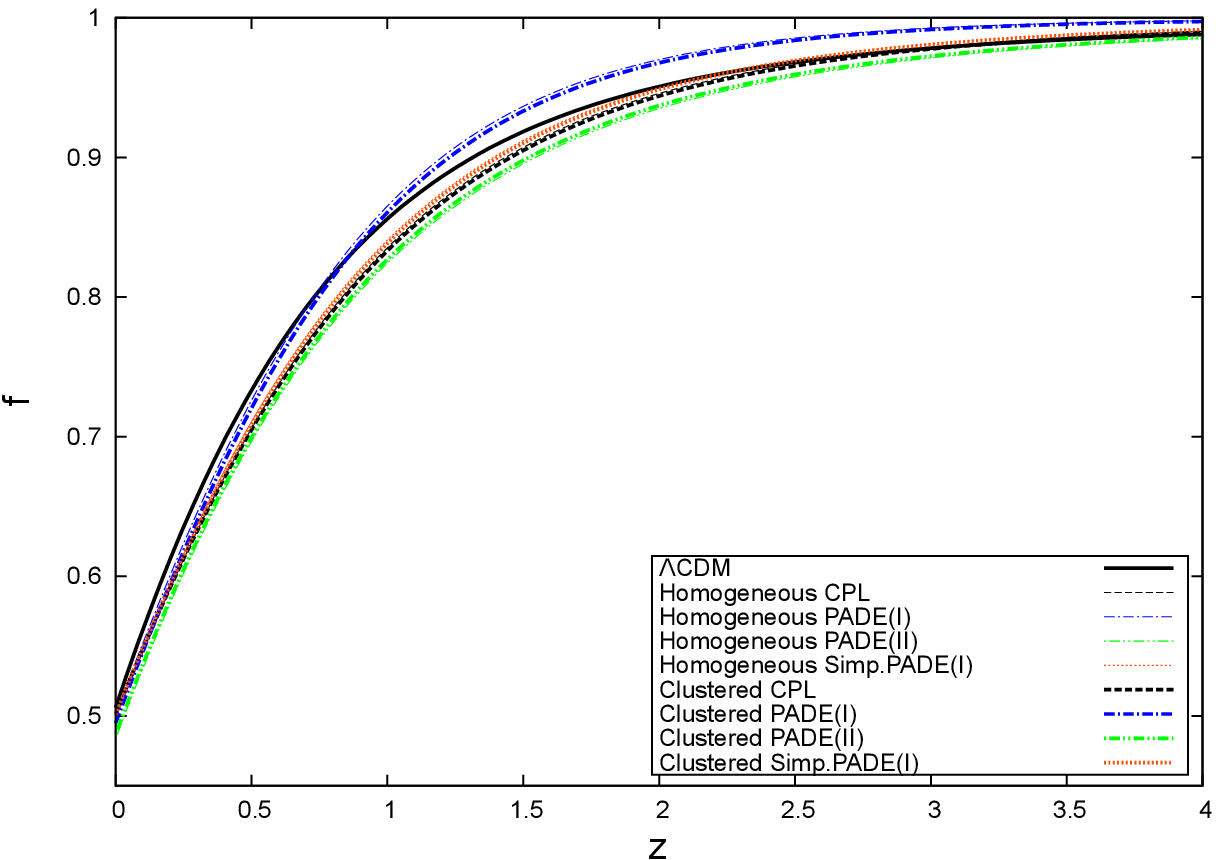}
	\includegraphics[width=8cm]{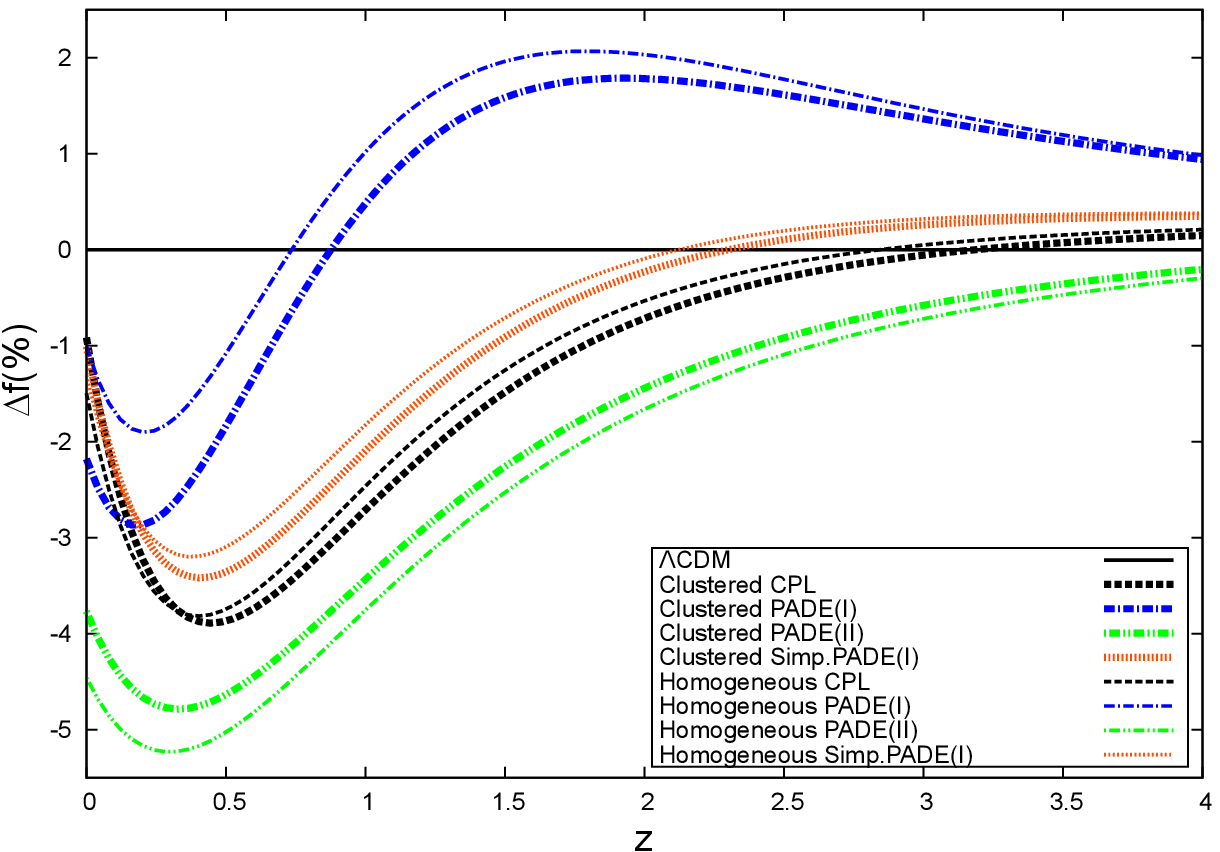}
	\caption{ The redshift evolution of the growth rate function $f(z)$ (top-panel) and the corresponding 
fractional difference $\Delta f(\%)=100\times [f(z)-f_{\rm \Lambda}(z)]/f_{\rm \Lambda}(z)$ (bottom panel).  
The different DE parameterisations are characterized by the colors and line-types presented
in the inner panels of the figure}.  
	\label{fig:growthfunction}
\end{figure}

\begin{figure} 
	\centering
	\includegraphics[width=8cm]{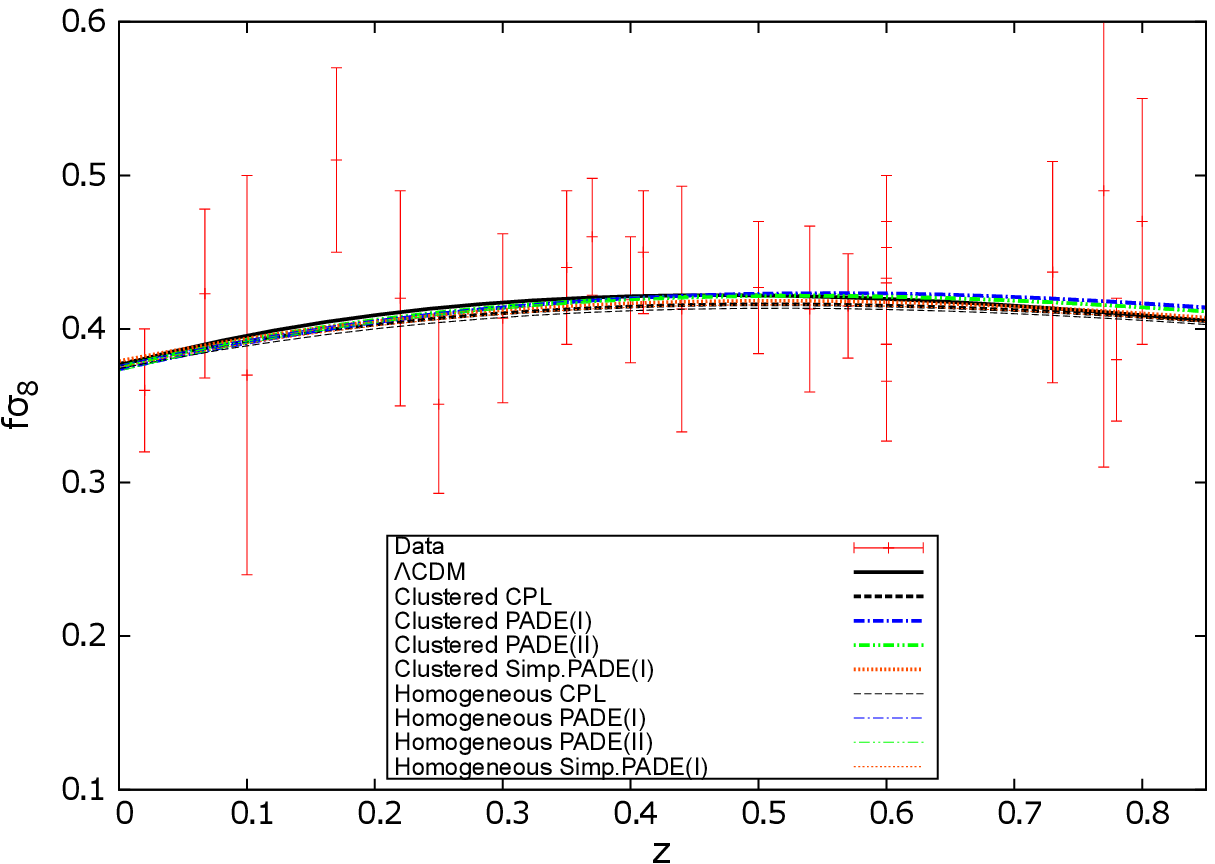}
	\caption{ Comparison of the observed and theoretical evolution of the
		growth rate $f (z)\sigma_8 (z)$ as a function of redshift $z$. Open squares correspond to the data.
		Line-types and colors are explained in the inner plot of the figure.}
	\label{fig:fsigma8}
\end{figure}
\section{Growth rate in DE parameterisations}\label{growth}
 In this section, we study the linear growth of matter perturbations in PADE cosmologies and we compare them with those of 
CPL and $\Lambda$CDM respectively.
In this kind of studies the natural question to ask is the following:
{\it how DE affects the linear growth of matter fluctuations?}
In order to treat to answer this question we need to introduce 
the following two distinct situations, which have
been considered within different approaches in the literature 
\citep{ArmendarizPicon:1999rj,Garriga:1999vw,ArmendarizPicon:2000dh,Erickson:2001bq,Bean:2003fb,Hu:2004yd,Abramo2007,Abramo2008,Ballesteros:2008qk,Abramo:2008ip,Basilakos:2009mz,dePutter:2010vy,Pace2010,Akhoury:2011hr,Sapone:2012nh,Pace2012,Batista:2013oca,Dossett:2013npa,Batista:2014uoa,Basse:2013zua,Pace2014a,Pace:2013pea,Pace:2014taa,Malekjani:2015pza,Naderi2015,Mehrabi:2014ema,Mehrabi:2015hva,Mehrabi:2015kta,Nazari-Pooya:2016bra,Malekjani:2016edh}: (i) the scenario 
in which the DE component is homogeneous ($\delta_{\rm de}\equiv 0$) and only the corresponding non-relativistic matter is allowed to cluster
($\delta_{\rm m}\ne 0$) and (ii) the case in which the whole system clusters 
(both matter and DE).
Owing to the fact that we are in the matter phase of the universe we 
can neglect the radiation term from the Hubble expansion.  

\subsection{Basic equations}
The basic equations that govern the evolution of non-relativistic matter and DE perturbations are given by \citep{Abramo:2008ip}
 \begin{eqnarray}
 &&\dot{\delta_{\rm m}}+\frac{\theta_{\rm m}}{a}=0\;,\label{grwoth1}\\
&& \dot{\delta_{\rm de}}+(1+w_{\rm de})\frac{\theta_{\rm de}}{a}+3H(c_{\rm eff} ^2 -w_{\rm de})\delta_{\rm de}=0\;,\label{grwoth2}\\
&& \dot{\theta_{\rm m}}+H \theta_{\rm m} - \frac{k^2 \phi}{a}=0\;,\label{grwoth3}\\
 &&\dot{\theta_{\rm de}}+H \theta_{\rm de} - \frac{k^2 c_{\rm eff} ^2 \theta_{\rm de}}{(1+w_{\rm de})a} - \frac{k^2 \phi}{a} =0\;,\label{grwoth4}
 \end{eqnarray}
where $k$ is the wave number and $c_{\rm eff}$ is the effective sound speed of perturbations
\citep{Abramo:2008ip,Batista:2013oca,Batista:2014uoa}. 
Combining the Poisson equation
  \begin{equation}
\label{Possnew}
 -\frac{k^2}{a^2}\phi=\frac{3}{2} H^2[\Omega_{\rm m} \delta_{\rm m} + (1+3 c_{\rm eff} ^2)\Omega_{\rm de} \delta_{\rm de}]\;,
\end{equation}
 with Eqs. (\ref{grwoth3} \& \ref{grwoth4}), eliminating $\theta_{\rm m}$ and $\theta_{\rm de}$ and changing the derivative 
from time to scale factor $a$, we obtain the following stystem
of differential equations \citep[see also][]{Mehrabi:2015kta,Malekjani:2016edh}
\begin{eqnarray}
 \delta_{\rm m}''+\frac{3}{2 a}(1-w_{\rm de} \Omega_{\rm de}) \delta_{\rm m}'=\frac{3}{2a^2}[\Omega_{\rm m} \delta_{\rm m} +\Omega_{\rm de}  (1+3 c_{\rm eff} ^2)\delta_{\rm de}]\;,\label{growth5}\\
 \delta_{\rm de}''+A \delta_{\rm de}'+B \delta_{\rm de}=\frac{3}{2a^2}(1+w_{\rm de})[\Omega_{\rm m} \delta_{\rm m} +\Omega_{\rm de}  (1+3 c_{\rm eff} ^2)\delta_{\rm de}]\;\label{growth6}.
 \end{eqnarray}
Bellow we set $c_{\rm eff}\equiv 0$ which means that the whole system (matter and DE) fully clusters. 
Moreover, we remind reader that for homogeneous DE models we have 
$\delta_{\rm de}\equiv 0$, hence Eq.(\ref{growth5}) reduces to the well known differential
equation of \cite{Peebles1993} [see also \cite{Pace2010} and references therein].
Concerning the functional forms of $A$ and $B$ we have
 \begin{eqnarray}
&&A=\frac{1}{a}[-3 w_{\rm de} -\frac{a w_{\rm de}'}{1+w_{\rm de}}+\frac{3}{2}(1-w_{\rm de} \Omega_{\rm de})],\nonumber \\  
&&B=\frac{1}{a^2}[-a w_{\rm de}' +\frac{a w_{\rm de}' w_{\rm de}}{1+w_{\rm de}}-\frac{1}{2}w_{\rm de}(1-3 w_{\rm de} \Omega_{\rm de})].
\end{eqnarray}
In order to perform the numerical integration of the above system (\ref{growth5} \& \ref{growth6}) 
it is crucial to introduce the appropriate initial conditions. Here we utilize 
\citep[see also][]{Batista:2013oca,Mehrabi:2015kta,Malekjani:2016edh}
 \begin{eqnarray}\label{initialcondition}
&& \delta_{\rm mi}'=\frac{\delta_{\rm mi}}{a_{\rm i}}\;,\nonumber \\  
&& \delta_{\rm dei}=\frac{1+w_{\rm dei}}{1-3w_{\rm dei}}\delta_{\rm mi}\;,\nonumber \\  
&&\delta_{\rm dei}'=\frac{4 w_{\rm dei}'}{(1-3w_{\rm dei})^2}\delta_{\rm mi}+\frac{1+w_{\rm dei}}{1-3w_{\rm dei}}\delta_{\rm mi}'\;,
\end{eqnarray}
where we fix $a_{\rm i}=10^{-4}$ and $\delta_{\rm mi}=1.5\times10^{-5}$.
In fact using the aforementioned conditions we verify that
matter perturbations always stay in the linear regime.
From the technical viewpoint, using $w_{\rm de}$, $\Omega_{\rm de}$ 
we can solve the system of equations (\ref{growth5} \& \ref{growth6}) 
which means that the fluctuations $(\delta_{\rm de},\delta_{\rm m})$ can be 
readily calculated, and from them $f(z)=d\ln{\delta_{\rm m}}/d\ln{a}$, 
$\sigma_8(z)=\frac{\delta_{\rm m}(z)}{\delta_{\rm m}(z=0)}\sigma_8(z=0)$ (rms mass variance at $R=8h^{-1}Mpc$) and $f(z)\sigma_8(z)$ immediately ensue.

Now we perform a joint statistical analysis
involving the expansion data (see Sect. \ref{back}) and the growth data.
In principle, this can help us to understand better the 
theoretical expectations of the present DE parameterisations, as well as to 
test their behaviour at the background and at the perturbation level.
The growth data and the details regarding the 
likelihood analysis ($\chi^{2}_{\rm gr}$, MCMC algorithm etc) can be found in Sect. \ref{back} of our 
previous work \citep{Mehrabi:2015kta}. 
Briefly, in order to obtain the overall likelihood function we need 
to include the likelihood function of the growth data in 
Eq.(\ref{eq:like-tot}) as follows
\begin{equation}
 {\cal L}_{\rm tot}({\bf p})={\cal L}_{\rm sn} \times {\cal L}_{\rm bao} \times {\cal L}_{\rm cmb} \times {\cal L}_{\rm h} \times
 {\cal L}_{\rm bbn}\times {\cal L}_{\rm gr}\;,
\end{equation}
hence 
\begin{equation}
\chi^2_{\rm tot}({\bf p})=\chi^2_{\rm sn}+\chi^2_{\rm bao}+\chi^2_{\rm cmb}+\chi^2_{\rm h}+\chi^2_{\rm bbn}+\chi^{2}_{\rm gr}\;,
\end{equation}
where the statistical vector ${\bf p}$ contains an 
additional free parameter, namely $\sigma_{8}\equiv \sigma_{8}(z=0)$.

In Tables (\ref{tab:best2}) and (\ref{tab:bestfit2}) we show the resulting
best fit values for various DE parameterisations under study, in which we also 
provide the observational constraints of the clustered DE parameterisations.
Furthermore, in Fig. (\ref{fig:contour_all}) we present 
the $1\sigma$ and  $2\sigma$ contours for 
various parameter pairs. The blue contour represents the confidence levels based on geometrical data and green ( red) contours show the confidence levels based on geometrical + growth rate data for  clustered (homogeneous) DE parameterisations.
Comparing the latter results with those of 
see Sect. \ref{back} we conclude that the observational constraints which are placed by 
the expansion+growth data are practically the same with those found by 
the expansion data. Therefore, we can use the current growth data in order 
to put constrains only on $\sigma_{8}$, since they do not significantly
affect the rest of the cosmological parameters.
This means that the results of see Sect. \ref{back}
concerning evolution of the main cosmological 
functions ($w_{\rm de}$, $E(z)$ and $\Omega_{\rm de}$) remain unaltered.
To this end, in 
Fig. (\ref{fig:growthfunction}) we plot the evolution of 
growth rate $f(z)$ as a function of redshift (upper panel) 
and the fractional difference with respect to that of $\Lambda$CMD model
(lower panel),
$\Delta f(\%)=100\times [f(z)-f_{\rm \Lambda}(z)]/f_{\rm \Lambda}(z)$. 
Specifically, in the range of $0\le z\le 4$ we find:
\begin{itemize}
  \item for homogeneous (or clustered) PADE I parameterisation the relative 
difference is 
$\sim [-1\%,1 \%]$ ( or $\sim [-2.25\%,1 \%]$)
\item in the case of simplified PADE I  we have 
$\sim [-1\%,0.4 \%]$ and $\sim [-1\%,0.4 \%]$ for homogeneous and clustered 
DE respectively
\item for homogeneous (or clustered) PADE II DE the relative 
deviation lies in the interval 
$\sim [-4.25\%,-0.25 \%]$ (or $\sim [-4\%,-0.25 \%]$).
Finally, in the case of CPL parameterisation
we obtain $\sim [-1.5\%, 0.1 \%]$ (homogeneous) and 
$\sim [-1\%, 0.1 \%]$ (clustered).
\end{itemize}

In this context, we verify that at high redshifts the growth rate tends 
to unity since the universe is matter dominated, namely 
$\delta_{\rm m}\propto a$. 
Moreover, we observe that 
the evolution of $\Delta f$ has one maximum/minimum and one zero point.
As expected, this feature of $\Delta f$ is related to the evolution of $\Delta E$ 
(see middle panel of Fig. \ref{fig:back}). Indeed, we verify that 
large values of the normalized Hubble parameter $E(z)$ correspond to 
small values of the growth rate. Also, looking at Fig. \ref{fig:back} ( middle panel) and 
Fig.\ref{fig:growthfunction} (bottom panel) we easily see that 
when $\Delta E$ has a maximum the growth rate $\Delta f$ has a minimum 
and vice versa.   
We also observe that if $\Delta E<0$ then $\Delta f >0$ and vice versa. 
Finally, in Fig. (\ref{fig:fsigma8}), we compare the observed 
$f\sigma_8(z)$ with the predicted growth rate function of the current
DE parameterisations [for curves see caption of Fig. (\ref{fig:fsigma8})].
We find that all parameterisations represent well the growth data.
As expected from AIC and BIC analysis (see Table \ref{tab:best2}) the current DE parameterisations and standard $\Lambda$CDM cosmology are all consistent with current observational
data. 

\subsection{The growth index}
We would like to finish this section with a discussion 
concerning the growth index of matter
fluctuations $\gamma$, which affects the growth rate of
clustering via the following relation 
\citep[first introduced by][]{Peebles1993} 
\be \label{fzz221} 
f(z)=\frac{d{\rm ln}
\delta_{\rm m}}{d{\rm ln} a}(z)\simeq \Omega^{\gamma}_{\rm m}(z) \;. 
\ee
The theoretical formula of the growth index has been studied
for various cosmological models,
including scalar field DE
\citep{Silveira:1994yq,Wang:1998gt,Linder:2003dr,Lue:2004rj,Linder:2007hg,Nesseris:2007pa},
DGP \citep{Linder:2007hg,Gong:2008fh,Wei:2008ig,Fu:2009nr},
Finsler-Randers \citep{Basilakos:2013ij}, running vacuum
$\Lambda(H)$ \citep{Basilakos:2015vra}, $f(R)$
\citep{Gannouji:2008wt,Tsujikawa:2009ku}, $f(T)$
\citep{Basilakos:2016xob}, Holographic DE
\citep{Mehrabi:2015kta} and Agegraphic DE \citep{Malekjani:2016edh}
If we combine equations
(\ref{grwoth1}-\ref{grwoth4}), (\ref{Possnew})
and using simultaneously $\frac{d\delta_{\rm m}}{dt}=aH\frac{d\delta_{\rm m}}{da}$
then we obtain \citep[see also][]{Abramo2007,Abramo:2008ip,Mehrabi:2015kta}
\begin{equation}\label{odedelta}
a^{2}\delta_{\rm m}^{\prime \prime}+
a\left(3+\frac{\dot{H}}{H^2}\right)\delta_{\rm m}^{\prime}=
\frac{3}{2}\Omega_{\rm m} \mu\;,
\end{equation}
where
\begin{equation}\label{eos22}
\frac{{\dot H}}{H^{2}}=\frac{d{\rm ln} H}{d{\rm ln}a}
=-\frac{3}{2}-\frac{3}{2}w_{\rm de}(a)\Omega_{\rm de}(a)\;,
\end{equation}
and $\Omega_{\rm de}(a)=1-\Omega_{\rm m}(a)$. The quantity 
$\mu(a)$ characterizes the nature of DE in PADE parametrisations, namely
\begin{equation} \label{VV}
\mu(a)=\left\{ \begin{array}{cc} 1
\;\;
       &\mbox{Homogeneous PADE}\\
  1+\frac{\Omega_{\rm de}(a)}{\Omega_{\rm m}(a)}\Delta_{\rm de}(a)(1+3c_{\rm eff}^2)
\;\;
       & \mbox{Clustered PADE}
       \end{array}
        \right.
\end{equation}
where we have set $\Delta_{\rm de}\equiv \delta_{\rm de}/\delta_{\rm m}$. 
Obviously, if we use $c_{\rm eff}^2=0$ then Eq.(\ref{odedelta})
reduces to Eq.(\ref{growth5}), while in the case of the usual $\Lambda$CDM
model we need to a priori set $\delta_{\rm de}\equiv 0$.

Furthermore, substituting Eq.(\ref{fzz221}) and Eq.(\ref{eos22}) in
Eq.(\ref{odedelta}) we arrive at 
\begin{equation}\label{Poll}
-(1+z)\frac{d\gamma}{dz}{\rm ln}(\Omega_{\rm m})+\Omega_{\rm m}^{\gamma}+
3w_{\rm de}\Omega_{\rm de}\left(\gamma-\frac{1}{2}\right)+\frac{1}{2}=
\frac{3}{2}\Omega_{\rm m}^{1-\gamma} \mu \;.
\end{equation}
Regarding the growth index evolution we use the following
phenomenological parameterisation \citep[see
also][]{Polarski:2007rr,Wu:2009zy,Belloso:2011ms,DiPorto:2011jr,Ishak:2009qs,Basilakos:2012ws,Basilakos:2012uu}
\begin{equation}
\label{gzzz}
\gamma(a)=\gamma_{0}+\gamma_{1}\left[1-a(z)\right]\;.
\end{equation}
Now, utilizing Eq.(\ref{Poll}) at the present time $z=0$ and
with the aid of Eq.(\ref{gzzz}) we obtain \citep[see
also][]{Polarski:2007rr}
\begin{equation}
\label{Poll2}
\gamma_{1}=\frac{\Omega_{\rm m0}^{\gamma_{0}}+3w_{\rm de0}(\gamma_{0}-\frac{1}{2})
\Omega_{\rm de0}+\frac{1}{2}-\frac{3}{2}\Omega_{\rm m0}^{1-\gamma_{0}} \mu_{0}}
{{\rm ln}  \Omega_{\rm m0} }\;,
\end{equation}
where $\mu_{0}=\mu(z=0)$ and $w_{\rm de0}=w_{\rm de}(z=0)$.
Clearly, in order to predict the growth index evolution in DE models 
we need to estimate the value of $\gamma_{0}$. 
For the current parameterisation it is easy to show that 
at high redshifts $z\gg 1$  
the asymptotic value of $\gamma(z)$ is written as 
$\gamma_{\infty}\simeq \gamma_{0}+\gamma_{1}$, while 
the theoretical formula of $\gamma_{\infty}$ is given by \cite{Steigerwald:2014ava}  
\be \label{g000}
\gamma_{\infty}=\frac{3(M_{0}+M_{1})-2(H_{1}+N_{1})}{2+2X_{1}+3M_{0}}
\ee where the following quantities have been defined: \be
\label{Coef1} M_{0}=\left. \mu \right|_{\rm \omega=0}\,, \ \
M_{1}=\left.\frac{d \mu}{d\omega}\right|_{\rm \omega=0} \ee and \be
\label{Coef2} N_{1}=0\,,\ \ H_{1}=-\frac{X_{1}}{2}=\frac{3}{2}\left.
w_{\rm de}(a)\right|_{\rm \omega=0} , 
\ee 
where $\omega={\rm ln}\Omega_{\rm m}(a)$.
Obviously, for $z\gg 1$ we get $\Omega_{\rm m}(a)\to
1$ [or $\Omega_{\rm de}(a)\to 0$] which implies $\omega \to 0$. 
For more details regarding the theoretical treatment of 
(\ref{g000}) we refer the reader to
\cite{Steigerwald:2014ava}. 
It is interesting to mention that the asymptotic value of the equation of state 
parameter for the current PADE cosmologies is written as 

\begin{eqnarray}
\label{w000}
w_{\infty}\equiv w_{\rm de}(a\to 0)=\left\{
\begin{array}{ll}
\frac{w_0+w_1}{1+w_2}\,,~~ & {\rm for}~~~{\rm PADE~ I}\,\\[4mm]
\frac{w_0}{1+w_{1}}\,, & {\rm for}~~~{\rm Sim.~PADE~ I} \\[4mm]
\frac{w_1}{w_2}\,, & {\rm for}~~~{\rm PADE~ II}\,.
\end{array} \right.
\end{eqnarray}

At this point we are ready to present our growth index results:
\begin{itemize}
  \item {\bf Homogeneous PADE parameterisations:} here we set
$\mu(a)=1$ ($\Delta_{\rm de}\equiv 0$). From 
Eqs.(\ref{Coef1}) and (\ref{Coef2}) we find
$$
\{ M_{0},M_{1},H_{1},X_{1}\}=\{ 1,0,\frac{3w_{\infty}}{2},-3w_{\infty}\}
$$
and thus Eq.(\ref{g000}) becomes 
\be 
\label{gmm}
\gamma_{\infty}
=\frac{3(w_{\infty}-1)}{6w_{\infty}-5} . 
\ee 
Lastly, inserting $\gamma_{0} \simeq \gamma_{\infty}-\gamma_{1}$ into
Eq.(\ref{Poll2}) and utilizing Eqs. (\ref{w000}-\ref{gmm}) 
together with 
the cosmological constraints of Table (\ref{tab:bestfit2})
we obtain 
\begin{eqnarray}
\label{gw000}
(\gamma_{0},\gamma_{1},\gamma_{\infty})=\left\{
\begin{array}{ll}
(0.555,-0.031,0.524)\,,~~ & {\rm for}~~~{\rm PADE ~I} \\[4mm]
(0.558,-0.021,0.537)\,, & {\rm for}~~~{\rm Sim.~PADE ~I} \\[4mm]
(0.559,-0.017,0.542)\,, & {\rm for}~~~{\rm PADE~II}\,.
\end{array} \right.
\end{eqnarray}
For comparison we provide the results for the $\Lambda$CDM model and CPL parameterisation
respectively. Specifically, we find 
$(\gamma_{0},\gamma_{1},\gamma_{\infty})_{\rm \Lambda}\simeq (0.556,-0.011,0.545)$ and 
$(\gamma_{0},\gamma_{1},\gamma_{\infty})_{\rm CPL}\simeq (0.561,-0.020,0.541)$. 

\item {\bf Clustered PADE parameterisations:} 
here the functional form of $\mu(a)$ is given
by the second branch of Eq.(\ref{VV}) which means that we need to define 
$\Delta_{\rm de}$. From Eq.(\ref{initialcondition}) we simply have 
$
\Delta_{\rm de}=\frac{1+{\rm w_{\rm de}}}{1-3{\rm w_{\rm de}}}$
and thus $\mu(a)$ takes the following form 
\be
\label{mm1}
\mu(a)=1+(1+3c_{\rm eff}^{2})\frac{\Omega_{\rm de}}{\Omega_{\rm m}}
\frac{(1+w_{\rm de})}{(1-3w_{\rm de})} \;. 
\ee 
In this case, from Eqs.(\ref{Coef1}) and (\ref{Coef2}) we obtain
(for more details see the Appendix)
$$
\{ M_{0},M_{1},H_{1},X_{1}\}=\{ 1,-\frac{(1+w_{\infty})(1+3c_{\rm eff}^2)}
{1-3w_{\infty}},\frac{3w_{\infty}}{2},-3w_{\infty}\}
$$
and from Eq.(\ref{g000}) we find

$$
\gamma_{\infty}\simeq \frac{3[(1-3w_{\infty})(1-w_{\infty})-(1+w_{\infty})(1+c_{\rm eff}^{2})]}{(6w_{\infty}-5)(3w_{\infty}-1)} \;.
$$
Notice, that in the case of fully clustered PADE parameterisations 
($c^{2}_{\rm eff}=0$) the above expression becomes
\be \label{g001}
\gamma_{\infty}\simeq \frac{3w_{\infty}(3w_{\infty}-5)}{(6w_{\infty}-5)(3w_{\infty}-1)} \;.
\ee
(\ref{w000}-\ref{gmm}) 
Now, utilizing Eqs.(\ref{w000}-\ref{g001}) and the
cosmological parameters of Table (\ref{tab:bestfit2}) we find 
\begin{eqnarray}
\label{gw000}
(\gamma_{0},\gamma_{1},\gamma_{\infty})=\left\{
\begin{array}{ll}
(0.547,0.005,0.552)\,,~~ & {\rm for}~~~{\rm PADE~I} \\[4mm]
(0.542,0.012,0.554)\,, & {\rm for}~~~{\rm Sim.~PADE~I} \\[4mm]
(0.549,0.003,0.552)\,, & {\rm for}~~~{\rm PADE~II}\,.
\end{array} \right.
\end{eqnarray}
To this end, 
if the CPL parameterisation is allowed to cluster then the asymptotic value
of the growth index is given by Eq.(\ref{g001}), where $w_{\infty}=w_{0}+w_{1}$.
In this case we find $(\gamma_{0},\gamma_{1},\gamma_{\infty})_{\rm CPL}\simeq (0.539,0.013,0.552)$. 

In Table (\ref{tab:index_gr}), we provide a compact presentation
of our numerical results including the
relative fractional difference $\Delta \gamma(\%)=[(\gamma-\gamma_{\rm \Lambda})/\gamma_{\rm \Lambda}]\times 100$ 
between all DE parameterisations and the 
concordance $\Lambda$ cosmology, in 3 distinct redshift bins.
Overall, we find that the fractional deviation lies in the interval 
$\sim [-2.2\%,0.3 \%]$. We believe that
relative differences of $|\Delta \gamma|\le 1\%$ 
will be difficult to detect even 
with the next generation of surveys, based mainly on
Euclid \citep[see][]{Taddei:2014wqa}. Using the latter forecast 
and the results presented in section \ref{growth}, we can now divide the current 
DE parameterisations into those that can be distinguished observationally
and those that are practically indistinguishable from $\Lambda$CDM model. 
The former DE parameterisations are the following three: homogeneous PADE I, 
clustered Simplified PADE I and clustered CPL. 
However, the reader has to remember
that these results are based on utilizing cosmological parameters that
have been fitted by the present day observational data (see Table \ref{tab:bestfit2}). 
Therefore, if future
observational data would provide slightly different values for the parameters of DE parameterisations 
then the growth rate predictions of the studied 
DE parameterisations could be somewhat
different than those derived here.

\begin{table*}
\tabcolsep 5pt 
\vspace {0.2cm}
\caption{Numerical results. The $1^{st}$ and the $2^{nd}$ 
	columns indicate the status of DE and the corresponding parametrisation.
	$3^{rd}$, $4^{th}$ and $5^{th}$
	columns show the $\gamma_{0}$, $\gamma_1$ and $\Delta_{\rm de0}$ values.
	The remaining columns present the fractional relative difference between the
	DE parameterisations and the $\Lambda$CDM cosmology based on the cosmological parameters 
	appeared in Table \ref{tab:bestfit2}.}
\begin{tabular}{lcccc|ccc|} \hline \hline
DE Status     & DE Parametrisation & $\gamma_{0}$ & $\gamma_1$ & $\Delta_{\rm de0}$& \multicolumn{3}{|c}{$\Delta \gamma$(\%)} \\
          &  &  &   &  &  $z<0.5$ & $0.5\le z <1$ & $1\le z <1.5$ \\ \hline
Homogeneous      & PADE I   & 0.555 & -0.031 & &-1.2 &  -1.8 &  -2.2 \\ 
                 & Sim. PADE I   & 0.558 & -0.021 & &-0.1&  -0.4 &  -0.6 \\ 
               & PADE II   & 0.559 & -0.017 & &0.15&  -0.01 &  -0.15 \\ 
               & CPL   & 0.561 & -0.02 & &0.3&  -0.02 &  -0.2 \\ 
Clustered      & PADE I   & 0.547 & 0.005 & 0.035&-0.8 &  -0.5 &  -0.1 \\ 
               & Sim. PADE I   & 0.542 & 0.012 & 0.047&-1.4 &  -0.7 &  -0.2 \\ 
               & PADE II   & 0.549 & 0.003 & 0.028&-0.6 &  -0.4 &  -0.02 \\ 
               & CPL  & 0.539 & 0.013 & 0.055&-2 &  -1.5 &  -0.8 \\ \hline 
\end{tabular}
\label{tab:index_gr}
\end{table*}

\end{itemize}

\section{Conclusions}
\label{conlusion}
We studied the cosmological properties of various DE parameterisations in which the EoS parameter is given with the aid of the 
PADE approximation. Specifically, using different types of
PADE parameterisation we investigated the behaviour of various DE scenarios
at the background and at the perturbation levels.

The main results of the present study are summarized as follows:

\textit{(i)} Initially, using the latest expansion data 
we performed a likelihood analysis in the context of Markov Chain Monte Carlo (MCMC) method.
It is interesting to mention that the statistical performance of the MCMC method has been discussed in \cite{Capozziello:2011tj} and references therein. Specifically, these authors showed that if we have  
	a multidimensional space of the cosmological parameters then the MCMC algorithm provides better constraints than other popular fitting procedures.
The results of our analysis for the explored PADE cosmologies, including those of CPL and $\Lambda$CDM 
can be found in Tables (\ref{tab:best} \& \ref{tab:bestfit}). Based on this analysis we placed constraints on the 
model parameters and we found that all DE parameterisations are consistent with the expansion data.
In this framework, using the best fit values we found that only the PADE (II) parameterisation remains in the quintessence regime ($−1 < w_{\rm de} < −1/3$). The rest of the 
PADE parameterisations evolves in the phantom region ($w_{\rm de}<-1$) at high redshifts, while 
they enter in the quintessence regime at relatively low redshifts. 
Concerning the cosmic expansion 
we found that prior to the present time the Hubble parameter of the DE parameterisations (PADE and CPL) is $\sim 2-3.5\%$ larger than the $\Lambda$CDM cosmological model.We also showed that the transition redshift from decelerating to accelerating expansion in the context of PADE parameterisations is consistent with that \citep{Farooq:2016zwm} using the cosmic chronometer $H(z)$ data. 
Notice, that similar results found in the framework of modified theory of gravities \citep{Capozziello:2014zda,Capozziello:2015rda}.  

\textit{(ii)} 
Then, we studied for the first time
the growth of perturbations in homogeneous and clustered PADE cosmologies.
First we used a joint statistical analysis
involving the expansion data and the growth data in order to place 
constraints on $\sigma_{8}$. Second,
based on the best fit cosmological parameters we computed the evolution of  
the growth rate of clustering $f(z)$. 
For the current DE parameterisations we found that the growth rate function 
is lower than $\Lambda$CDM model at low redshifts, 
while the differences among the parameterisations are negligible at high redshifts. 
Third, following the notations of \cite{Steigerwald:2014ava} we derived the functional form 
of the growth index ($\gamma$) of linear matter perturbations. 
Assuming that DE is homogeneous
we found the well known asymptotic value of the growth index, namely  
$\gamma_{\infty}=\frac{3(w_{\infty}-1)}{6w_{\infty}-5}$ [$w_{\infty}=w(z\to \infty)$], while 
in the case of clustered DE parameterisations we obtained 
$\gamma_{\infty}\simeq \frac{3w_{\infty}(3w_{\infty}-5)}{(6w_{\infty}-5)(3w_{\infty}-1)}$.

Finally, utilizing the fractional deviation between all DE parameterisations and the 
concordance $\Lambda$ cosmology we found that $\Delta \gamma \sim [-2.2\%,0.3 \%]$. We concluded 
that relative differences of $|\Delta \gamma|\le 1\%$ 
will be difficult to detect even 
with the next generation of surveys, based on
Euclid \citep[see][]{Taddei:2014wqa}. Combining the latter forecast 
and the results presented in section \ref{growth}, we divided the current DE 
parameterisations into those that can be distinguished observationally
and those that are practically indistinguishable from $\Lambda$CDM. 
The former DE parameterisations are the following three: homogeneous PADE I, 
clustered Simplified PADE I and clustered CPL. 

\section{Acknowledgements}
MM and AM acknowledge support from the Iran Science Elites Federation.
DFM acknowledges support from the Research Council of Norway, and the NOTUR facilities.
SB acknowledges support by the Research Center 
for Astronomy of the Academy of Athens in the
context of the program ''Testing general relativity on cosmological scales''
(ref. number 200/872).

\appendix
\section{$M_{1}$ coefficient for clustered dark energy models} 
Here we provide some details 
concerning the coefficient $M_{1}$ which 
appears in Eq.(\ref{g000}). This 
coefficient is given in terms of 
the variable $\omega={\rm \ln}\Omega_{\rm m}$ (see section 4.1)
which means that when $a\to 0$ ($z\gg 1$) we get $\Omega_{\rm m}\to 1$ 
(or $\omega \to 0$). 
From Eq.(\ref{Coef1}) we have
$$
M_{1}=\left.\frac{d \mu}{d\omega}\right|_{\omega=0}=
\left.\Omega_{\rm m}\frac{d \mu}{d\Omega_{\rm m}}\right|_{\Omega_{\rm m}=1} .
$$
Of course, in the case of homogeneous dark energy, namely $\mu(a)=1$ 
we simply find $M_{1}=0$. However, if dark energy 
is allowed to cluster then 
the situation becomes quite different. 

Indeed, using Eq.(\ref{mm1})
we 
obtain after some calculations 
\begin{eqnarray*}
\frac{d \mu}{d\Omega_{\rm m}}=
(1+3c^{2}_{\rm eff})\frac{d }{d\Omega_{\rm m}}\left( \frac{\Omega_{\rm de}}{\Omega_{\rm m}}\right)
\frac{1+w_{\rm de}}{1-3w_{\rm de}}+(1+3c^{2}_{\rm eff})\nonumber\\
\times\frac{\Omega_{\rm de}}{\Omega_{\rm m}}
\frac{d }{d\Omega_{\rm m}}\left( \frac{1+w_{\rm de}}{1-3w_{\rm de}}\right)
\end{eqnarray*} 
where $\Omega_{\rm de}=1-\Omega_{\rm m}$, 
$$
\frac{d }{d\Omega_{\rm m}}\left( \frac{\Omega_{\rm de}}{\Omega_{\rm m}}\right)=-\frac{1}{\Omega^{2}_{\rm m}},
$$
$$
\frac{d }{d\Omega_{\rm m}}\left( \frac{1+w_{\rm de}}{1-3w_{\rm de}}\right)=
\frac{d }{da}\left( \frac{1+w_{\rm de}}{1-3w_{\rm de}}\right)\frac{da}{d\Omega_{\rm m}}
$$
with 
$$
\frac{d\Omega_{\rm m}}{da}=\frac{3}{a}\Omega_{\rm m}(1-\Omega_{\rm m})w_{\rm de}=
\frac{3}{a}\Omega_{\rm m}\Omega_{\rm de}w_{\rm de}.
$$
Obviously, based on the above equations we arrive at
\begin{eqnarray*} 
\Omega_{\rm m}\frac{d \mu}{d\Omega_{\rm m}}=
-(1+3c^{2}_{\rm eff})
\frac{1+w_{\rm de}}{\Omega_{\rm m}(1-3w_{\rm de})}+(1+3c^{2}_{\rm eff})\frac{a}{3\Omega_{\rm m}w_{\rm de}}\\
\times\frac{d }{da}\left( \frac{1+w_{\rm de}}{1-3w_{\rm de}}\right)
\end{eqnarray*}
Taking the limit $\Omega_{\rm m}\to 1$ ($a\to 0$) of the latter expression
we calculate $M_{1}$ 
$$
M_{1}=\left.\Omega_{\rm m}\frac{d \mu}{d\Omega_{\rm m}}\right|_{\Omega_{\rm m}=1}=
-(1+3c^{2}_{\rm eff})\frac{1+w_{\rm de}}{1-3w_{\rm de}}.
$$

\bibliographystyle{aasjournal}
\bibliography{ref}
\label{lastpage}

\end{document}